\documentclass[twocolumn]{aastex631}

\usepackage{hyperref}
\usepackage{CJKutf8}
\usepackage{physics}
\usepackage{enumitem}
\usepackage{xcolor}

\begin{document}

\title{Closing the stellar labels gap: Stellar label independent evidence for [$\alpha$/M] information in \emph{Gaia} BP/RP spectra}
\shortauthors{Laroche \& Speagle}
\shorttitle{Closing the stellar labels gap}

\author[0000-0002-5522-0217]{Alexander Laroche}
\affiliation{David A. Dunlap Department of Astronomy \& Astrophysics, University of Toronto, 50 St George St, Toronto, ON M5S 3H4, Canada}
\affiliation{Dunlap Institute for Astronomy \& Astrophysics, University of Toronto, 50 St George Street, Toronto, ON M5S 3H4, Canada}

\author[0000-0003-2573-9832]{Joshua S. Speagle (\begin{CJK*}{UTF8}{gbsn}沈佳士\end{CJK*})}
\affiliation{Department of Statistical Sciences, University of Toronto, 9th Floor, Ontario Power Building, 700 University Ave, Toronto, ON M5G 1Z5, Canada}
\affiliation{David A. Dunlap Department of Astronomy \& Astrophysics, University of Toronto, 50 St George St, Toronto, ON M5S 3H4, Canada}
\affiliation{Dunlap Institute for Astronomy \& Astrophysics, University of Toronto, 50 St George Street, Toronto, ON M5S 3H4, Canada}
\affiliation{Data Sciences Institute, University of Toronto, 17th Floor, Ontario Power Building, 700 University Ave, Toronto, ON M5G 1Z5, Canada}

\begin{abstract}
Data-driven models for stellar spectra which depend on stellar labels suffer from label systematics which decrease model performance: the “stellar labels gap”. \textcolor{black}{To close the stellar labels gap, we present a stellar label independent model for \emph{Gaia} BP/RP spectra. We develop a novel implementation of a variational auto-encoder, which learns to generate an XP spectrum and accompanying ‘scatter’ without relying on stellar labels.} We demonstrate that our model achieves competitive XP spectra reconstructions in comparison to stellar label dependent models. We find that our model learns stellar properties directly from the data itself. We then apply our model to XP/APOGEE giant stars to study the [$\alpha$/M] information in \emph{Gaia} XP. We provide strong evidence that the XP spectra contain meaningful [$\alpha$/M] information by demonstrating that our model learns the $\alpha$-bimodality, without relying on stellar label correlations for stars with $T_{\rm eff} <$ 5000 K, \textcolor{black}{while also being sensitive to the anomalous abundances of \emph{Gaia}-Enceladus stars.} We publicly release our trained model, codebase and data. Importantly, our stellar label independent model can be implemented for any/all XP spectra because our model performance scales with training object density, not training label density.
\end{abstract}

\section{Introduction} \label{sec:intro}

The advent of increasingly large-scale spectroscopic surveys, such as APOGEE \citep{APOGEE-17}, LAMOST \citep{LAMOST-22}, GALAH \citep{GALAH-21}, and most recently \emph{Gaia} BP/RP \citep[XP;][]{GDR3-22}, has motivated astronomers to develop data-driven models to cope with the massive influx of data \citep[e.g.][]{Cannon-15,Payne-19,Cycle-StarNet-21,AstroNN-19,Zhang++23,leung-LAM-23,ASPGAP-23}. These data-driven techniques often seek to discern stellar properties from stellar spectra and/or generate stellar spectra from stellar properties.

Typically, data-driven methods are differentiated from physics-driven methods in the following way: A model which relies on synthetic stellar spectra is physics-driven, in the sense that it explicitly attempts to match theoretical spectra to observations. Conversely, a model is data-driven if it does not rely on theoretical spectra. Generally, data-driven models incorporate some sort of machine learning algorithm to shed theoretical modeling. 

Data-driven models have begun to be favoured over their physics-driven predecessors due to a disconnect between theory and practice known as the “synthetic gap.” \textcolor{black}{The synthetic gap is a combination of theoretical systematics and instrumental effects which in tandem produce discrepancies between synthetic and observed stellar spectra, including but not limited to: one-dimensional modeling, assumption of hydrostatic equilibrium and local thermal equilibrium, and telluric lines. Although significant progress has been achieved in stellar atmosphere modeling, e.g. three-dimensional non-local thermal equilibrium models \citep{Bergemann14}, these simplified modeling assumptions are still frequently adopted when determining stellar properties of large-scale surveys.}  For a more comprehensive overview, see the Introduction of \cite{Cycle-StarNet-21}.

However, it is seldom emphasized that any data-driven model which relies on stellar labels, a term which typically refers to effective temperature $T_{\rm eff},$ surface gravity $\log g,$ metallicity [M/H] (and occasionally $\alpha$-abundance $[\alpha$/M]), is \emph{implicitly} physics-driven. \textcolor{black}{Indeed, stellar labels utilized during training of such a data-driven model are typically estimated from theoretical stellar spectra. This is not strictly the case: e.g. $T_{\rm eff}$ and $\log g$ for \emph{Gaia} FGK benchmark stars were determined with angular diameter measurements and bolometric fluxes \citep{Heiter++15}. However, such samples contain nowhere near the requisite number of stars for training, although useful for validation (34 \emph{Gaia} FGK benchmark stars). } Furthermore, this ab initio physics-driven estimation is often times several generations removed from the data-driven model being implemented, as it is becoming increasingly more common to ‘train machine learning on machine learning.’ \textcolor{black}{In this context: training a machine learning model on stellar labels which were themselves obtained from a machine learning model. It has been demonstrated that training large language models on synthetic data produces a decrease in output diversity which worsens with each successive ‘ML on ML’ iteration \citep{Guo+23}. A similar concern for stellar properties is therefore warranted.} 

We therefore introduce the concept of an additional gap: the “stellar labels gap,” ecompassing stellar label systematics which negatively impact performance of data-driven models which rely on stellar labels (stellar label dependent models). The stellar labels gap includes:
\begin{enumerate}[noitemsep]
    \item[(i)] Poorly estimated stellar labels
    \item[(ii)] Regions of stellar label space where labels are insufficient summary statistics for a spectrum
    \item[(iii)] Regions of stellar label space with a dearth of labels to train on
    \item[(iv)] Stellar multiplicity\footnote{While in principle stellar label models can account for binarity, in practice this is rarely implemented.}
    (binaries, triples, etc.)
\end{enumerate}
These stellar label systematics then lead to systemic bias in stellar label dependent model predictions. In order to close the stellar labels gap, astronomers should develop purely data-driven models which do not rely on stellar labels (stellar label independent models).

\textcolor{black}{To that end, this work presents a stellar label independent model for \emph{Gaia} XP spectra: an unsupervised learning model which applies data compression.}  We develop a novel implementation of a \textcolor{black}{variational auto-encoder (VAE); a \emph{scatter} VAE}, which learns to generate an XP spectrum while simultaneously estimating intrinsic scatter for individual XP spectra. We demonstrate certain advantages of our stellar label independent approach by comparing our model performance to stellar label dependent models. Subsequently, we interpret the behaviour of model by contrasting stellar label space to our latent space. We then apply our model to the high- and low-$\alpha$ sequences to provide stellar label independent evidence that the \emph{Gaia} XP spectra contain meaningful [$\alpha$/M] information.

The subsequent Sections of this paper are organized as follows: Section \ref{sec:data} briefly reviews the data used in this work, namely \emph{Gaia} XP spectra and APOGEE stellar labels. Section \ref{sec:meth} presents our stellar label independent model architecture as well as our model training procedure. Section \ref{sec:res} compares our trained model performance to stellar label dependent models and interprets what our model has learned about the XP spectra. We then use our model, in Section \ref{sec:app} to conclusively demonstrate that the \emph{Gaia} XP spectra contain meaningful [$\alpha$/M] information, without the well known issue of $\alpha$-abundance correlations with stellar labels, for stars with $T_{\rm eff} <$ 5000 K. Finally, in Section \ref{sec:conc} we conclude by discussing the implications of our work in the context of stellar label in/dependent modeling of stellar spectra, its limitations, and promising future applications.

\section{Data}\label{sec:data}

\begin{figure*}
    \centering
    \includegraphics[width=\textwidth]{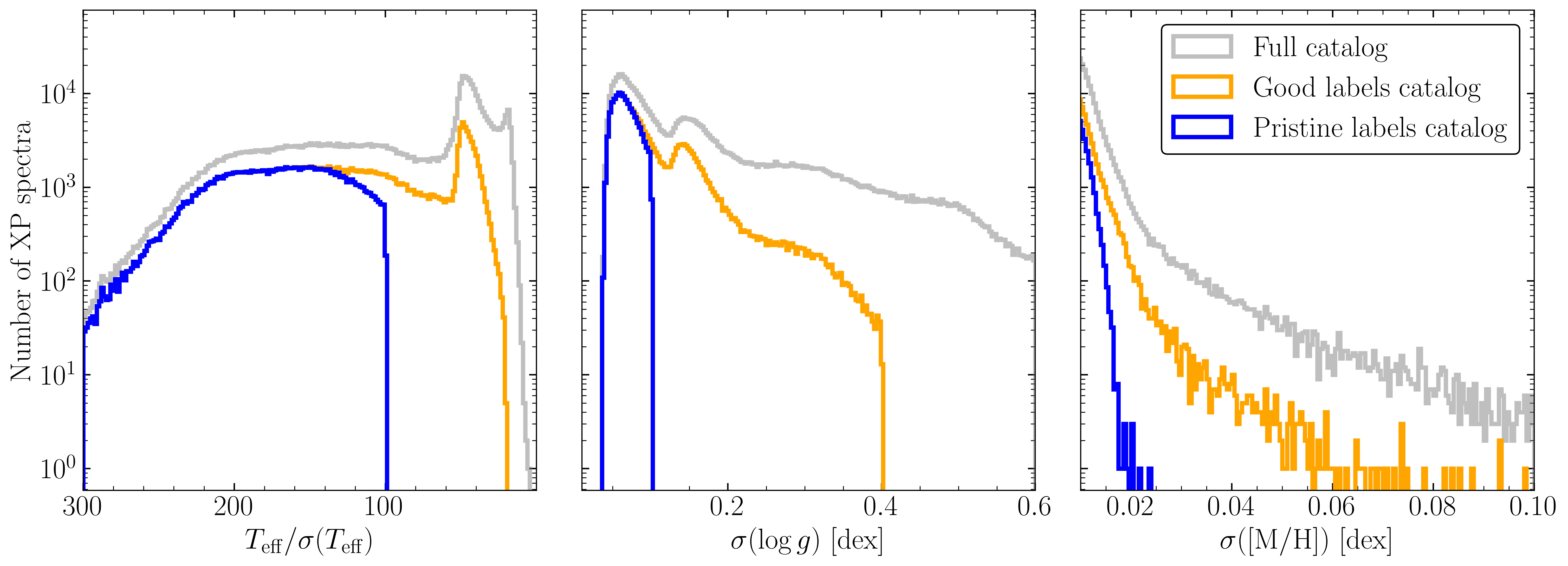}
    \caption{Signal-to-noise ratio/error distributions over stellar labels for the full, good labels and pristine labels catalogs defined in Section \ref{ssec:cat}. $T_{\rm eff}$ SNR, $\log g$ uncertainty and [M/H] uncertainty from the \texttt{Astro-NN} value-added APOGEE catalog are presented is left, middle and right panels, respectively. From the full, to the good labels, to the pristine labels catalog, quality cuts become increasingly restrictive. The full catalog is used to train our model, the good labels catalog is used interpret our model behaviour, and the pristine labels catalog is used to investigate $\alpha$-abundance information in the XP spectra.}
    \label{fig:cuts}
\end{figure*}

In this Section, we review the data used in this work: the \emph{Gaia} XP spectra, as well as the APOGEE derived stellar labels we use in order to $(i)$ compare our stellar label independent model to stellar label dependent models and $(ii)$ interpret the behaviour of our model.

\subsection{Gaia low-resolution BP/RP spectra}

The \emph{Gaia} low-resolution XP spectra in \emph{Gaia} Data Release 3 \citep[GDR3,][]{GDR3-22} is comprised of 220+ million flux-calibrated, low-resolution spectra \citep{GDR3-XP-22,GDR3-instrument-22}. These spectra combine measurements from the Blue Photometer (BP) and Red Photometer (RP) \textit{Gaia} instruments which span 330-680 and 640-1050 nm, respectively. 

We use the XP coefficient spectra in this work. XP spectra deviate from traditional spectra because they are typically reported in Hermite polynomial space. Specifically, the BP and RP spectra are transformed from \textit{discrete} wavelength-space into \textit{continuous} coefficient-space: 110 coefficients which weight a set of Hermite polynomials (55 blue and 55 red coefficients). One can transform coefficients into fluxes, and vice-versa, ideally without loss of information. Previous machine learning models have been successfully implemented when using data in both XP wavelength space and coefficient space. For instance, \citet{leung-LAM-23} train their large astronomy model in XP coefficient space, whereas \citet{Zhang++23} train their deep stellar label model in XP wavelength-space. 

\subsection{Catalogs}\label{ssec:cat}

To construct our train/test datasets, we perform a cross-match betweeen the GDR3 XP spectra and the \texttt{astroNN}\footnote{https://github.com/henrysky/astroNN} value-added stellar label catalog for APOGEE Data Release 17 \citep[DR17,][]{AstroNN-19,APOGEE-22}. In principle, our model can be trained on the entire XP dataset, since our model does not require stellar labels. However, a primary focus of this work is to compare our stellar label independent model to stellar label dependent models, as well as compare generative spaces: stellar label space versus latent space. As such, we restrict ourselves to the cross-matched XP/APOGEE dataset. For the purposes of this work, we construct three different catalogs from the XP/APOGEE cross-match: the full catalog, the good labels catalog and the pristine labels catalog:

\begin{itemize}[noitemsep]
    \item The \emph{full catalog} is the complete XP/APOGEE cross-match with no quality cuts whatsoever, which contains 502,311 stars. This is the dataset we use to train our model. Stellar label dependent models almost exclusively require quality cuts on training labels, which inherently limits the amount of training data. However, our stellar label independent model can train on XP spectra which have poor APOGEE stellar labels.

    \item The \emph{good labels catalog} is the XP/APOGEE cross-match restricted to APOGEE labels with high signal-to-noise ratio (SNR), which contains 202,970 stars (approximately 40\% of the full catalog). We obtain the good labels catalog with the following quality cuts:
    \begin{enumerate}[label=(\roman*),noitemsep]
        \item $T_{\rm eff}/\sigma(T_{\rm eff})>30$
        \item $\sigma(\log g)<0.4$ 
        \item $\sigma([\rm M/H])<0.2$ 
        \item $0<BP-RP<4$
        \item $6<G<17.5$
        \item No bit set in \texttt{STARFLAG}
        \item No bit 19 (\texttt{M\_H\_BAD}) or bit 23 (\texttt{STAR\_BAD}) in \texttt{ASPCAPFLAG}
    \end{enumerate}

    \item The \emph{pristine labels catalog} is created from analogous cuts to the good labels catalog, except the stellar label cuts are even more restrictive:
    \begin{enumerate}[label=(\roman*),noitemsep]
        \item $T_{\rm eff}/\sigma(T_{\rm eff})>100$
        \item $\sigma(\log g)<0.1$ 
        \item $\sigma([M/H])<0.05$ 
     \end{enumerate}
    The pristine stellar label cuts, in combination with the same color, magnitude and flag cuts from the good labels catalog, yield 123,804 stars (approximately 25\% of the full catalog).
    
\end{itemize}
We depict the stellar label cuts which define the full, good labels and pristine labels catalogs in Figure \ref{fig:cuts}. For the good labels catalog, the effective temperature (surface gravity, metallicity) cuts remove 16,028 (43,705, 32,314) stars. For the pristine labels catalog, the effective temperature (surface gravity, metallicity) cuts remove 273,466 (286,191, 33,769) stars.

\begin{figure*}[htb!]
    \centering
    \includegraphics[width=\textwidth]{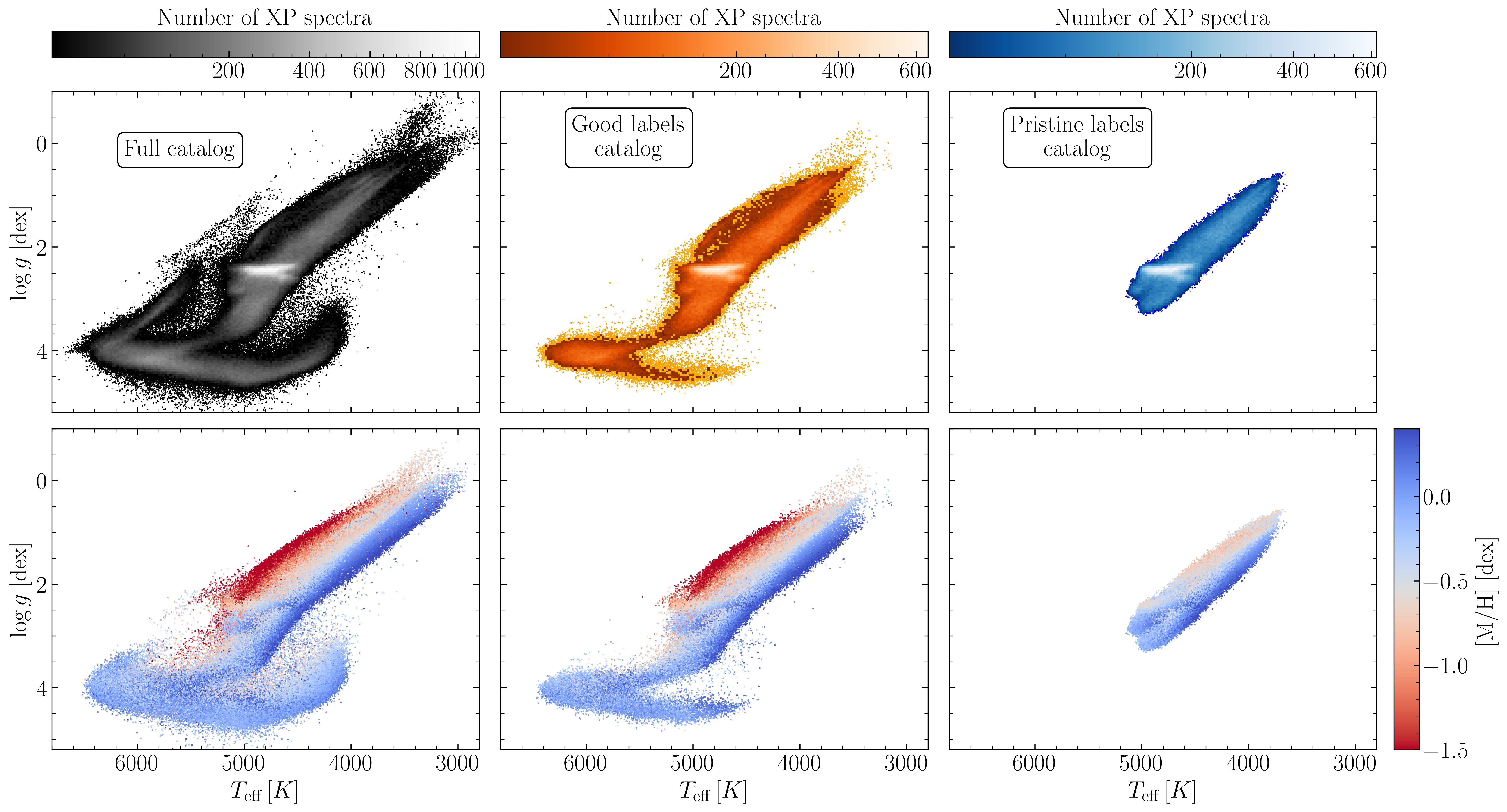}
    \caption{Kiel diagrams for the full (left), good labels (middle) and pristine labels (left) catalogs, colored by XP spectra number density (top row) and metallicity (bottom row). The increasingly restrictive quality cuts for the good labels and pristine labels catalog relative to the full catalog shrink the stellar label space they cover.}
    \label{fig:kiel}
\end{figure*}

We compare the Kiel diagrams of the three catalogs in Figure \ref{fig:kiel}. \textcolor{black}{The full catalog appears to have poorly estimated labels at both ends of the main sequence, where surface gravities are systematically overestimated, and the metallicity gradient along the giant branch is somewhat obfuscated.} Conversely, the good labels catalog has both a ‘cleaner’ main sequence and ‘sharper’ giant branch metallicity gradient. The pristine catalog is then exclusively composed of giant stars with very accurate stellar labels. \textcolor{black}{Finally, apart from the stellar label cuts defining the pristine catalog removing all main-sequence stars, we do not observe major differences between the apparent magnitude and \emph{Gaia} color distributions across our catalogs, beyond the color and magnitude cuts we impose.}

\subsection{Preprocessing}\label{meth:pre}

Before training our model on the full catalog, we perform the following pre-processing on the XP coefficient spectra. We first adopt the now common practice of concatenating the 55 BP and 55 RP coefficients into a single vector of length 110. Second, we normalize each individual spectrum by its corresponding \emph{Gaia} $G-$band mean flux. Finally, we standard normalize the XP spectra coefficient by coefficient to zero mean and unit variance. Recently, \cite{Zhang++23} implemented median/quantile normalization for XP preprocessing to reduce the negative impact of outliers during training. However, one of the strengths of our model is the ability to incorporate outliers through our star-by-star scatter estimation, described in Section \ref{meth:svae}. As such, we opt for standard normalization. 

\section{Method: stellar label independent generative model}\label{sec:meth}

This work presents a stellar label independent model which generates XP spectra. Specifically, we develop a novel implementation of a variational auto-encoder (VAE): \emph{scatter} variational auto-encoder (\emph{s}VAE). Before describing our \emph{s}VAE architecture, we briefly review the concept of a (V)AE.

\subsection{Variational auto-encoder review}

An AE, which \textcolor{black}{needs} not be variational, can be thought of as a non-linear generalization of Principal Component Analysis. An AE begins with an encoder which compresses input data, in our case a stellar spectrum, to a low-dimensional latent representation. A decoder then attempts to reconstruct the stellar spectrum from the latent representation. Ideally, this will allow the latent representation to learn key features which are shared across a set of stellar spectra. The variational nature of a VAE is added to an AE by upgrading the latent space from a collection of discrete points to a latent distribution $\mathcal{Z}$. The most popular VAE methodology is that of \citet{VAE-13}, who encode input data onto an independent multivariate Gaussian distribution. The latent space can therefore be entirely characterized by a latent mean vector $\mu$ and variance vector $\sigma',$ with latent space dimension $n_{\mathcal{Z}}.$

\subsection{Scatter variational auto-encoder}\label{meth:svae}

\begin{figure*}[htb!]
    \centering
    \includegraphics[width=\textwidth]{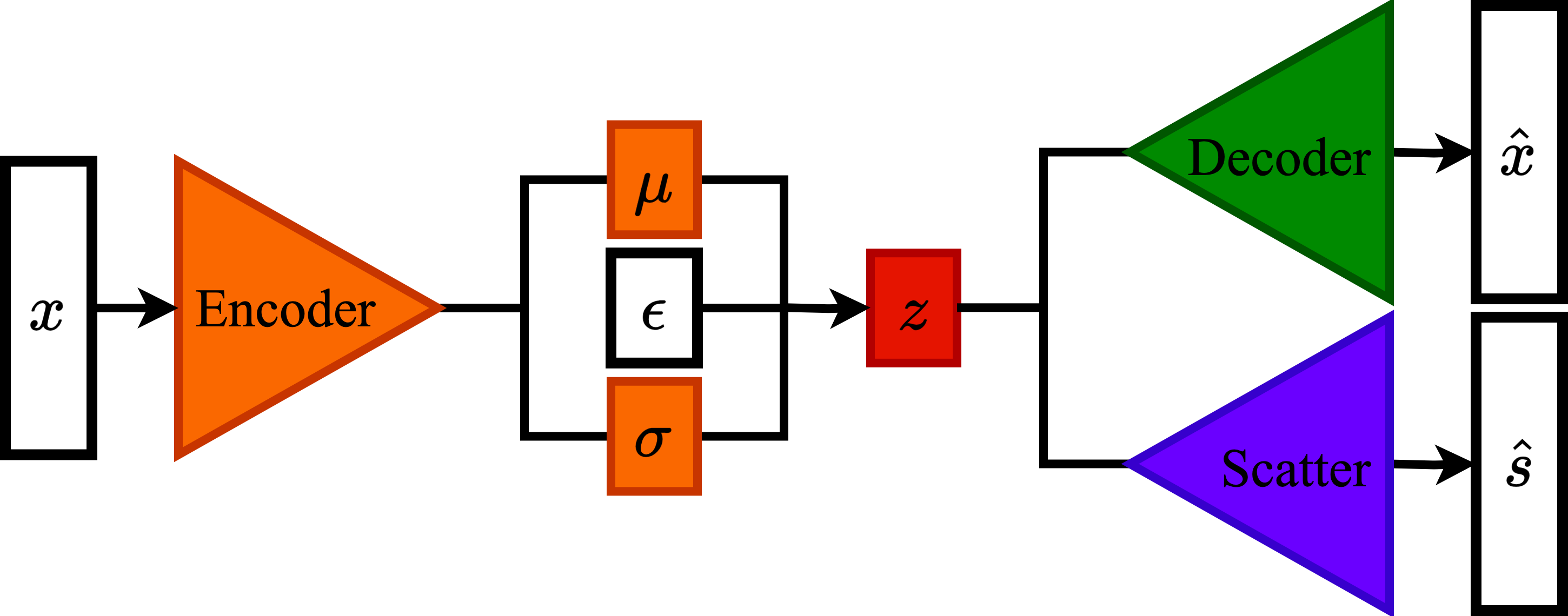}
    \caption{The \emph{scatter} VAE architecture: An XP spectrum $x$ is compressed into latent means $\mu$ and latent variances $\sigma.$ The re-parametrization trick through $\epsilon$ then projects the encoding into the latent space $z.$ Subsequently, the decoder reconstructs an estimate of the XP spectrum $\hat{x},$ and the scatter decoder produces a scatter estimate for the spectrum $\hat{s}.$ During training, $x$ and $\hat{x}$ are encouraged to be as similar as possible, while simultaneously enforcing Gaussian structure in the latent space.}
    \label{fig:svae_arch}
\end{figure*}

We present the high-level architecture of our model in Figure \ref{fig:svae_arch}. Our \emph{s}VAE differs from a traditional VAE since, in addition to a decoder (green) which estimates stellar spectra, we introduce a second ‘decoder’ which estimates intrinsic scatter \emph{on a star-by-star basis} (purple). Importantly, both the XP reconstruction and XP scatter estimate are generated from the same latent space (red). After discarding the encoder (orange) post-training, new XP coefficients (and scatter) can be generated given an arbitrary latent space vector.

The input of our encoder is a pre-processed XP spectrum. The XP coefficients are fed through five intermediate layers, composed of 90, 70, 50, 30 and 10 neurons, respectively. All intermediate layers are activated by the gaussian error linear unit (GELU). The latent parameters are then given by a linear transformation of the final intermediate layer. In this work, we have fixed the latent space to 6 dimensions, meaning the encoder produces 12 outputs (6 means and variances). \textcolor{black}{This choice was informed by initial experimentation with latent dimensions ranging from 1 to 20 dimensions. We found that a 6 dimensional latent space produced an optimal balance between interpretability and reconstruction error.} Note that as is typical with VAEs, the reparametrization trick randomly samples $\epsilon\sim\mathcal{N}(0,1)$ (standard normal distribution) to transform the latent mean $\mu$ and variance $\sigma$ vectors into a latent space vector via $z=\epsilon\mu+\sigma.$ Our decoder is the mirror image of our encoder, and takes as input a vector drawn from the latent distribution. We then reconstruct an XP spectrum, by feeding the latent vector through five intermediate layers analogous to the encoder, except in reverse order. Finally, a reconstruction of the 110 XP coefficients is produced with a linear transform. The scatter estimator has the same architecture as the decoder, except we enforce positivity with a final Sigmoid activation. Importantly, weights and biases of the scatter estimator are entirely disconnected from the decoder. 

\textcolor{black}{The interpretation of what the scatter estimator represents is by no means straightforward. The scatter estimator does not produce an estimate of intrinsic scatter in the traditional sense: variance of the entire XP dataset assuming zero measurement error. Rather, the scatter could be both ‘intrinsic’ to an individual star: stellar variability arising from stellar spots, pulsations, etc. or simply be anomalous relative to the training set, but also ‘extrinsic’: arising from either $(i)$ information loss through our data compression procedure or $(ii)$ \emph{Gaia} systematics, poor observations or underestimation of uncertainties. As such, it is more accurate to think of the scatter estimate as an error term which includes traditional intrinsic scatter, systematics, outliers, etc. Empirically, we noticed a marked increase in performance when including the scatter estimator in our model. Therefore, despite not being able to pinpoint precisely which of the above aspects the scatter estimate mitigates, we can assert that our \emph{s}VAE architecture does improve spectral modeling.}

\begin{figure*}
    \centering
    \includegraphics[width=\textwidth]{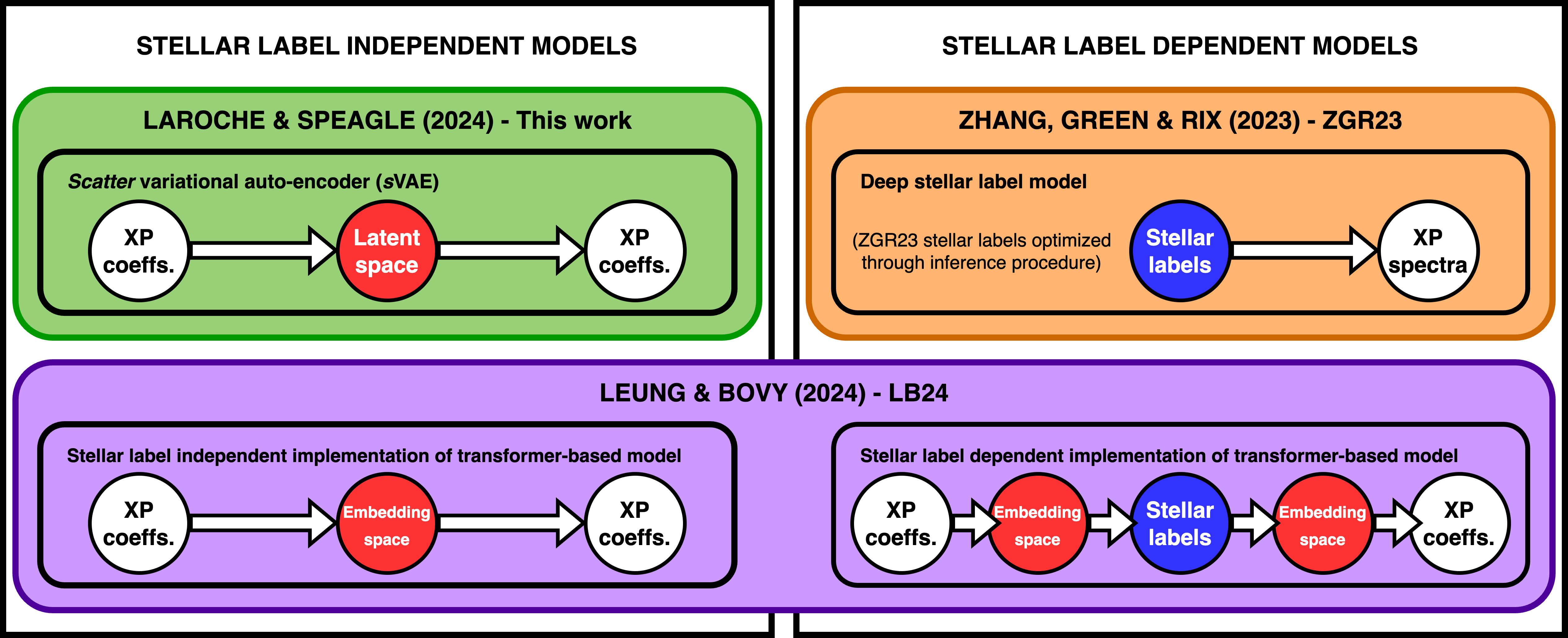}
    \caption{A high-level overview of the models used in this work. We classify stellar independent models (\emph{left}) and stellar dependent models (\emph{right}) by whether or not stellar labels are used to represent an XP spectrum. Our \emph{s}VAE (green) projects an XP coefficient spectrum into the latent space, and subsequently reconstructs the coefficients. The stellar label independent implementation of LB23 (purple, \emph{left}) projects an XP coefficient spectrum into the embedding space, and subsequently reconstructs the coefficients. On the other hand, ZGR23 (orange) generates an XP spectrum, in wavelength space, from stellar labels which were optimized through their inference procedure. Finally, the stellar label dependent implementation of LB23 (purple, \emph{right}) projects an XP coefficient spectrum into the embedding space, estimates stellar labels, then re-projects the labels into the embedding space to ultimately reconstruct the coefficients.}
    \label{fig:models}
\end{figure*}

\subsection{Training}

\textcolor{black}{After data pre-processing (see Section \ref{meth:pre}), we randomly split the full catalog into 90\% for training, and allocate the remaining 10\% for test data.} We adopt a similar training approach to \cite{leung-LAM-23}, and use the \texttt{AdamW} optimizer \citep{Adam-14,AdamW-17} with Cosine Annealing with Warm Restarts \citep{warm-cos-anneal-16} as our learning-rate scheduler. Cosine Annealing is a non-monotic learning rate \textcolor{black}{scheduler} with cosine-shaped cycles. We select an initial cycle learning rate of $10^{-4}$ and a final learning rate of $10^{-10}.$ We train our \emph{s}VAE for 5000 epochs, with 10 cycles of 500 epochs. As in \cite{leung-LAM-23}, we find that a batch size of 1024 is ideal for balancing training speed and model performance. 

We denote the $j^{\rm th}$ coefficient of the the $i^{\rm th}$ XP spectrum as $x_{ij},$ where $j=1,\dots,110$ with accompanying uncertainties $\sigma_{ij},$ and $i$ spans the size of our training set. Furthermore, we denote the \emph{s}VAE estimate of a coefficient (from the decoder) for a given XP spectrum as $\hat{x}_{ij},$ with accompanying scatter $\hat{s}_{ij}$ (from the scatter estimator). During training, we aim to minimize the following loss function to optimize our \emph{s}VAE model parameters:
\begin{align}\label{eq:loss}
    \mathcal{L} = 
    \Tilde{\chi}^2(x,\hat{x},\sigma,\hat{s}) + D_{\rm KL}(\mu, \sigma').
\end{align}
$\Tilde{\chi}^2$ is the reconstruction loss between the input XP spectrum $x$ and the reconstructed spectrum $\hat{x}$, whilst incorporating observational uncertainties $\sigma$ and intrinsic scatter $\hat{s}.$ In Eq. \eqref{eq:loss}, the $\tilde{\chi}^2$ term is given by
\begin{align}\label{eq:chi2}
    \Tilde{\chi}^2 = \chi^2(x,\hat{x},\sigma,\hat{s}) + P(\sigma,\hat{s}).
\end{align}
The first term in Eq. \eqref{eq:chi2} is the traditional (reduced) $\chi^2,$ given by
\begin{align}
    \chi^2(x,\hat{x},\sigma,\hat{s})
    = \frac{1}{110N}\sum_{i=1}^N\sum_{j=1}^{110}
    \frac{(x_{ij}-\hat{x}_{ij})^2}{\sigma_{ij}^2+\hat{s}_{ij}^2},
\end{align}
and the second is a penalty term $P$ given by
\begin{align}
    P(\sigma,\hat{s})
    = \frac{1}{110N}\sum_{i=1}^N\sum_{j=1}^{110}
    \log\left(\sigma_{ij}^2+\hat{s}_{ij}^2\right),
\end{align}
The second term in Eq. \eqref{eq:loss} is the latent space structure loss, for which we select the KL divergence \citep{KLD-51}, and is given by
\begin{align}\label{eq:KLD}
    D_{\rm KL}(\mu, \sigma') = 
    \frac{1}{6N}\sum_{i=1}^{N}\sum_{j=1}^{6}
    \left[\mu_{ij}^2 + \sigma_{ij}^{'2} - \left(1+\log\sigma_{ij}^{'2}\right)\right],
\end{align}
where $\mu$ ($\sigma'$) are the latent means (variances) and we are summing over the 6 latent space dimensions. In the above form, $D_{\rm KL}$ is not particularly intuitive. For the purposes of latent space structure loss, it can be thought as a distance between probability distributions. $D_{\rm KL}=0$ if two distributions are identical and $D_{\rm KL}>0$ otherwise. Hence, assuming a multivariate normal latent distribution, the KL divergence is a measure of the Gaussianity of the \emph{s}VAE latent space. \textcolor{black}{Note that no $z$ variables directly appear in Eq. \eqref{eq:KLD} due to marginalizing out $z$ from the joint probability distribution $P(z|\mu,\sigma').$} 

\textcolor{black}{Finally, we emphasize that Eq. \eqref{eq:loss} does not account for the full \emph{Gaia} XP coefficient covariance matrix. Our decision to neglect covariances during training is based on computational feasibility. Specifically, the inclusion of the full covariance matrix would have lead to an increased computational cost due to both loss function evaluation and the scatter output necessarily being augmented to two dimensions (to match the covariance matrix) if we took this approach.}

\subsection{Data and code}

Our \emph{s}VAE model, implemented in \texttt{PyTorch}, can be trained on a single NVIDIA RTX4070 Ti GPU in $\sim$6 hours, with our current XP/APOGEE cross-match of $\sim450,000$ training objects\footnote{Linearly extrapolating this training time to the full \emph{Gaia} DR3 XP dataset of 220 million spectra, we roughly estimate that training on all currently available XP spectra would take $\sim$ 120 GPU days (on the same GPU, with the same batch size, architecture, etc.). To give some perspective, a foundation language model such as \texttt{LLaMA} \citep{touvron2023llamaopenefficientfoundation} takes 21 days $\times$ 2048 GPUs = $\sim43,000$ GPU days to train.}. Our model can project $\sim10^5$ XP spectra into the latent space, and simulate XP spectra from the latent space in a few seconds. Our trained stellar label independent model and codebase are publicly available at \href{https://github.com/AlexLaroche7/xp_vae/releases/tag/v1.0.0}{\texttt{https://github.com/AlexLaroche7/xp\_vae}} for others to reproduce our results, build upon our existing model and apply our model to XP spectra beyond the XP/APOGEE cross-match. Furthermore, all data associated with this work is available at \href{https://zenodo.org/records/10951393}{\texttt{https://zenodo.org/records/10951393}}.

\begin{figure*}
    \centering
    \includegraphics[width=\textwidth]{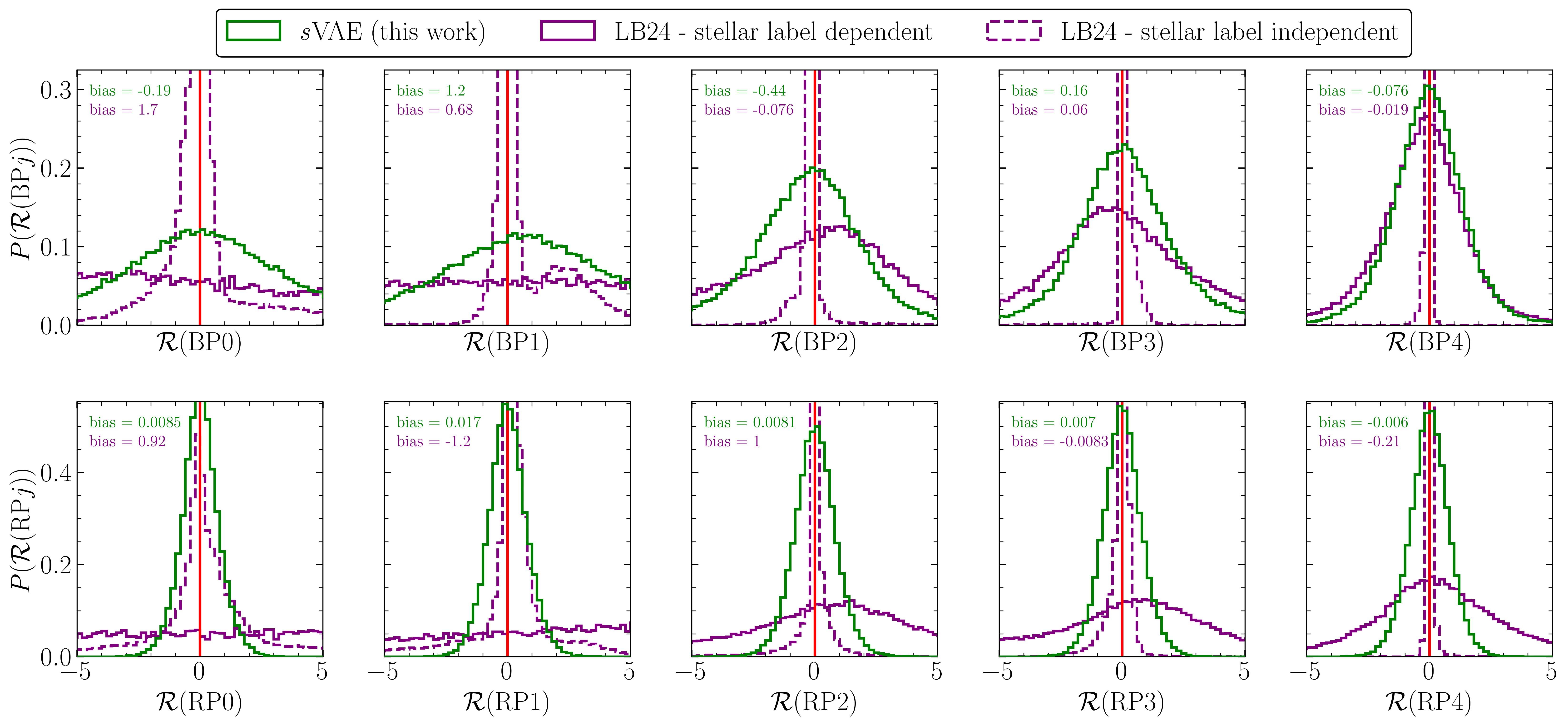}
    \caption{Relative coefficient reconstruction error distributions, $\mathcal{R}(j)$ from Eq. \eqref{eq:coeff_rel_err}, for the stellar label dependent (solid purple) and independent (dashed purple) implementations of \cite{leung-LAM-23} in comparison to our stellar label independent model (green), over test data in the full catalog for the first 5 BP and RP coefficients. Bias is presented for each coefficient for our \emph{s}VAE and the stellar label dependent LB23 model. Our \emph{s}VAE outperforms the stellar label dependent LB23 model, but underperforms relative to the stellar label independent LB23 model, which suffers from less information loss due to its larger embedding space (64 tokens in comparison to 6 latent variables).}
    \label{fig:lam_cat_coeff_errs}
\end{figure*}
\begin{figure}
    \centering
    \includegraphics[width=\columnwidth]{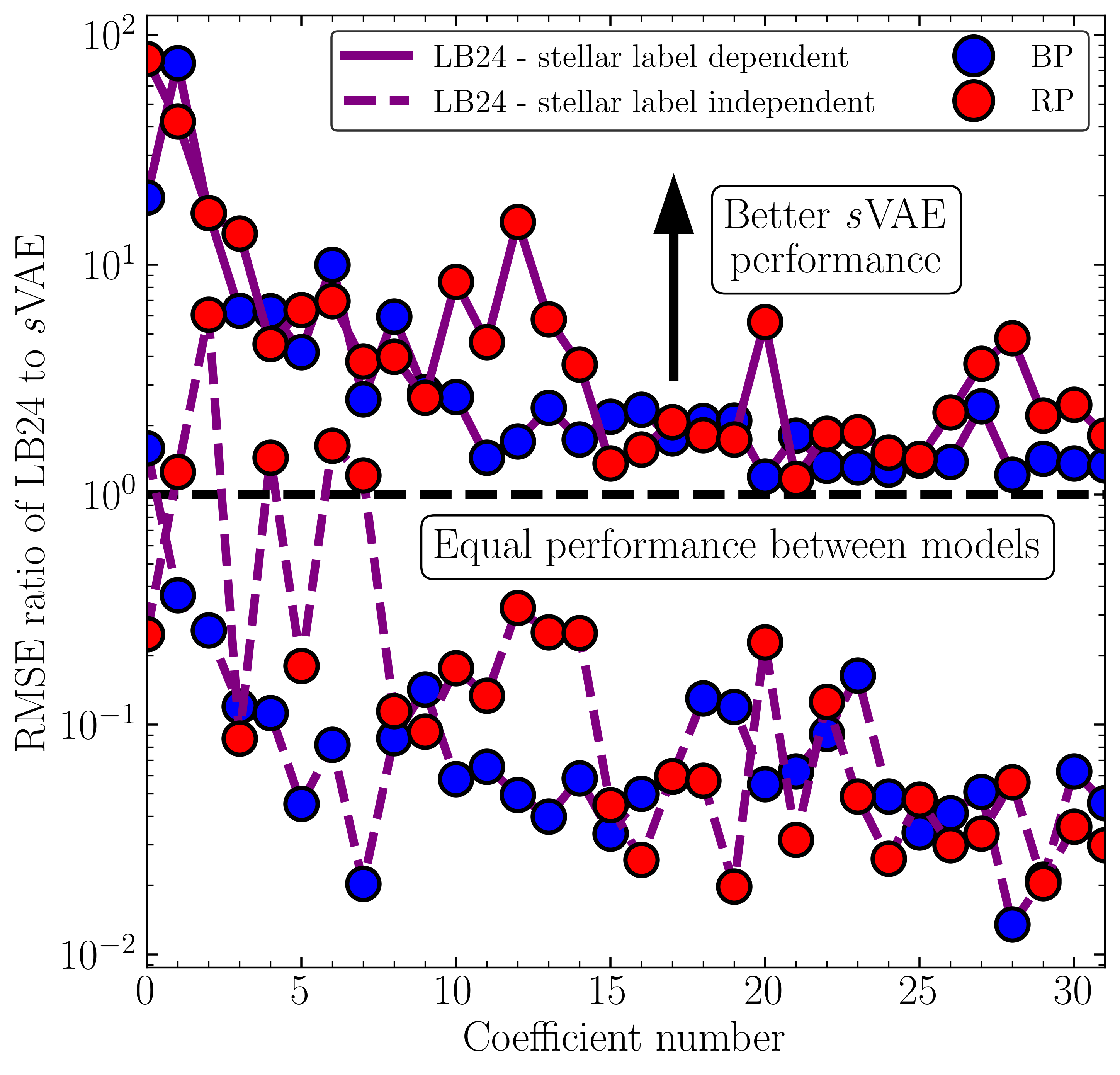}
    \caption{Ratio of spectrum reconstruction error as a function of BP/RP coefficient, RMSE$(j)$ from Eq. \eqref{eq:rmse_coeff}, for the stellar label independent and dependent implementations of LB23 relative to our stellar label independent model. Our \emph{s}VAE produces more accurate reconstructions of the lower order coefficients, and decays to negligible improvement for noise-dominated higher order coefficients, relative to the stellar label dependent LB23 model. Conversely, the stellar label independent LB23 model largely outperforms our \emph{s}VAE due to less information loss.}
    \label{fig:compare_lam}
\end{figure}

\section{Results}\label{sec:res}

\begin{figure*}
    \centering
    \includegraphics[width=\textwidth]{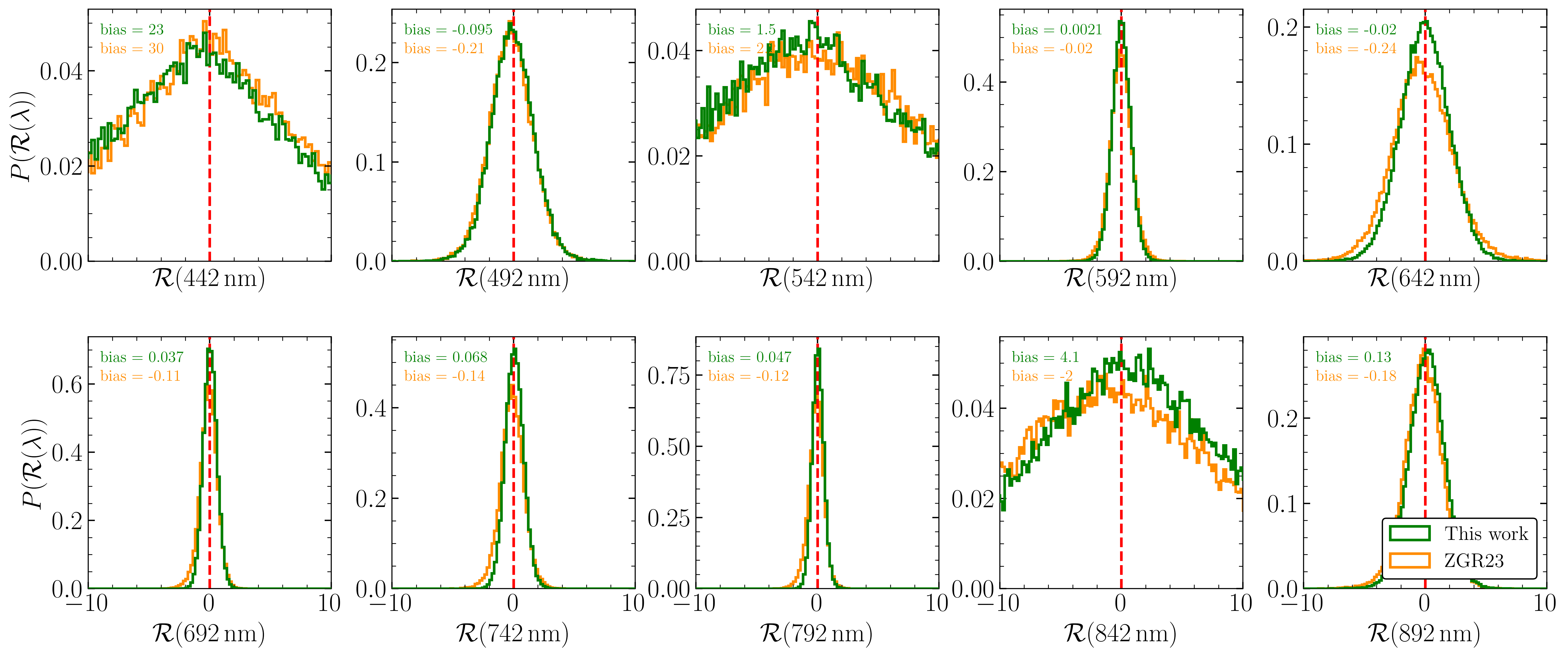}
    \caption{Relative wavelength reconstruction error distributions, $\mathcal{R}(\lambda)$ from Eq. \eqref{eq:lam_rel_err}, for the deep stellar label model of \cite{Zhang++23} (ZGR23, \textcolor{black}{orange}) in comparison to our stellar label independent model (green), over test data in the full catalog for 10 wavelengths uniformly distributed across the XP wavelengths. Bias is presented at each wavelength. Here, we apply the reliability cut of ZGR23 to both sets of reconstruction errors. From 600-800 nm, our \emph{s}VAE produces less bias than ZGR23.}
    \label{fig:zgr_cat_wl_errs}
\end{figure*}

\begin{figure}
    \centering
    \includegraphics[width=\columnwidth]{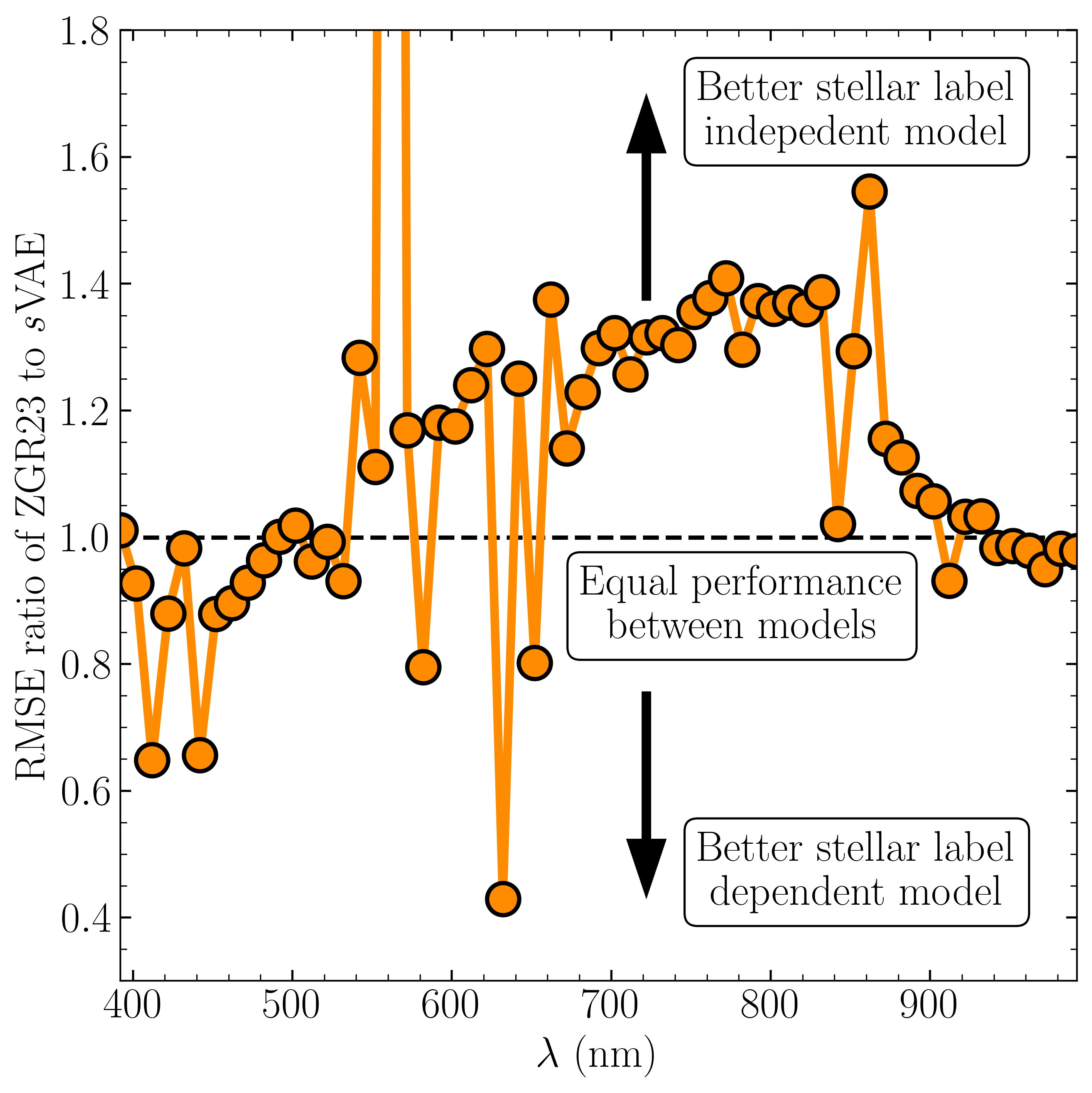}
    \caption{Ratio of spectrum reconstruction error as a function of BP/RP wavelength, RMSE$(\lambda)$ from Eq. \eqref{eq:rmse_lam}, for ZGR23 relative to our stellar label independent model. Similar to Figure \ref{fig:zgr_cat_wl_errs}, we apply the ZGR23 reliability cut to both sets of reconstruction errors. From 550-900 nm, our \emph{s}VAE produces more accurate wavelength reconstructions than ZGR23 over most wavelengths, by 20-40\%.}
    \label{fig:compare_zgr}
\end{figure}

We now present the results of our trained \emph{s}VAE. First, we compare our model performance to two models from the literature in Section \ref{res:perf}. Second, we demonstrate the behaviour of our model by observing stellar label trends in the \emph{s}VAE latent space in Section \ref{res:latent}.

\subsection{Model performance}\label{res:perf}

To compare the performance of our stellar label independent approach to stellar label dependent models, we compare our model to two generative models for XP spectra (see Figure \ref{fig:models}):
\begin{enumerate}
    \item\textit{Large astronomy model} \citep[][LB23]{leung-LAM-23}: A transformer-based model which transfers the methods of Large Language Models (LLMs) to astronomical data: a ‘large astronomy model’ (LAM). LB23 can perceive/predict any combination of XP data or stellar labels. As such, we compare with LB23 by implementing their model in both a stellar independent \emph{and} dependent way. Specifically:
    \begin{enumerate}[label=(\roman*)]
        \item \emph{Stellar label independent}: We project XP coefficients into the embedding space, and subsequently reconstruct the coefficients.
        \item \emph{Stellar label dependent}: We project the XP coefficients into the embedding space, estimate stellar labels from the embedding space, re-project the labels into the embedding space, and finally reconstruct the coefficients. The stellar labels for LB23 are: $T_{\rm eff}$, $\log g$ and [M/H] and $J-H$ and $J-K$ colours from 2MASS photometry \citep{2MASS_06}. Near-infrared photometry is included to provide a proxy for extinction by breaking the temperature-extinction degeneracy.
    \end{enumerate}
    Note that the LB23 model has a context window length of 64: the maximum number of tokens their model can handle. As such, we only project/reconstruct the first 32 BP and RP coefficients (64 total) with their model.
    
    \item\textit{Deep stellar label model} \citep[][ZGR23]{Zhang++23}: A data-driven model which estimates stellar parameters. ZGR23 can also generate XP spectra from $T_{\rm eff}$, $\log g,$ [M/H], extinction and distance. Note that because ZGR23 produce their own stellar parameter estimates, we use the ZGR23 stellar parameters catalog to generate XP spectra with their model. Additionally, ZGR23 advise that their results are only reliable for stars which pass their reliability cut, which we enforce by requiring \texttt{quality\_flags} $<$ 8 (see Section \ref{sec:res} for further discussion on the implications of their reliability cut).
\end{enumerate}
We compare our model reconstruction errors over test data (10\% of the full catalog) to errors from both stellar label independent models. LB23 and ZGR23 are similar in the sense that they both generate XP spectra from labels, but dissimilar in their output: LB23 predicts XP spectra in coefficient space, whereas ZGR23 predicts XP spectra in wavelength space. In order to compare to ZGR23, we convert our model XP coefficient predictions to wavelength space with \texttt{GaiaXPy}.\footnote{\url{https://gaia-dpci.github.io/GaiaXPy-website/}, version 2.1.0: \url{10.5281/zenodo.8239995} \citep{daniela_ruz_mieres_2023_8239995}}

To compare our model to LB23, we also define the \emph{relative coefficient reconstruction error} $\mathcal{R}(j)$ as the relative error for a single coefficient for a single star:
\begin{equation}\label{eq:coeff_rel_err}
    \mathcal{R}(j) \equiv 
    \frac{x_{ij}-\hat{x}_{ij}}{\sigma_{ij}},
\end{equation}
where $\mathcal{R}\equiv\mathcal{R}(j)$ because the error is computed in XP coefficient space, over $j.$ \textcolor{black}{We also define the
\emph{spectrum reconstruction error} $\mathrm{RMSE}(j)$ as the root mean square error over coefficients for a single star:
\begin{equation}\label{eq:rmse_coeff}
\begin{aligned}
    \mathrm{RMSE}^2(j) &\equiv 
    \frac{1}{64}\left\{(x_{BP}-\hat{x}_{BP})^TC_{BP}^{-1}(x_{BP}-\hat{x}_{BP})\right. \\
    &+ \left.(x_{RP}-\hat{x}_{RP})^TC_{RP}^{-1}(x_{RP}-\hat{x}_{RP})\right\},
\end{aligned}
\end{equation}
where $\mathrm{RMSE}(j)$ computes the reconstruction error while accounting for the XP covariance matrices. Specifically, $x_{BP}$ and $x_{RP}$ are vectors containing the first 32 BP and RP coefficients due to the length 64 context window of LB23. Similarly, $\hat{x}_{BP}$ and $\hat{x}_{RP}$ are the model reconstructions for the first 32 BP and RP coefficients, and $C_{BP}$ ($C_{RP}$) are the 32$\times$32 sub-matrices of the full BP (RP) covariance matrices.}

Next, to compare our model to ZGR23, we define analogous quantities to the former two errors, but in XP wavelength space. We define the \emph{relative wavelength reconstruction error} $\mathcal{R}(\lambda)$ as the relative error for a single wavelength for a single star:
\begin{equation}\label{eq:lam_rel_err}
    \mathcal{R}(\lambda) = \frac{x_i(\lambda)-\hat{x}_i(\lambda)}{\sigma_i(\lambda)},
\end{equation}
where $\mathcal{R}\equiv\mathcal{R}(\lambda)$ because the error is computed in wavelength space, over $\lambda.$ Lastly, we define the \emph{spectrum reconstruction error} $\mathrm{RMSE}(\lambda)$ as the root mean square error over wavelengths for a single star:
\begin{equation}\label{eq:rmse_lam}
    \mathrm{RMSE}(\lambda) \equiv 
    \sqrt{\frac{1}{61}\sum_{j=1}^{61}
    \mathcal{R}^2(\lambda)},
\end{equation}
where $\mathrm{RMSE}(\lambda)$ is integrating the square of
$\mathcal{R}(\lambda)$ over wavelengths. Note that $j\in[1,61]$ because ZGR23 predict the flux at 61 XP wavelengths\footnote{\textcolor{black}{\texttt{GaiaXPy} does not currently have the functionality to account for the full XP covariance matrices, and as such we only consider univariate errors in Eq. \eqref{eq:rmse_lam}.}}. 

In Figure \ref{fig:lam_cat_coeff_errs}, we compare the relative coefficient error distributions for our model to LB23 across the first five BP and RP coefficients, Eq. \eqref{eq:coeff_rel_err}. Our relative error distributions are Gaussian with negligible bias, relative to the stellar label dependent LB23 model. Conversely, the stellar label independent LB23 model evidently outperforms our model over most coefficients. Additionally, in Figure \ref{fig:compare_lam}, we compare the ratio of coefficient reconstruction errors over the full catalog for LB23 relative to our trained \emph{s}VAE, Eq. \eqref{eq:rmse_coeff}, for both the stellar label independent and dependent implementations. We find that, for the lowest order XP coefficients, our model outperforms the stellar label dependent LB23 model by 1-2 orders of magnitude. Furthermore, the relative performance increase for our model decays to unity for higher order coefficients, which are the most noise dominated. Conversely, we observe that the stellar label independent LB23 model outperforms our \emph{s}VAE by an order of magnitude for all but the lowest order coefficients. 

We conclude that our \emph{s}VAE outperforms the stellar label dependent model, but not the stellar label independent model, of LB23. This is expected, given the length 64 context window of LB23. In other words, the stellar label independent LB23 model is effectively learning a one-to-one mapping between coefficients, without data compression. \textcolor{black}{In contrast, our model compresses the XP spectrum into 6 latent variables, and as such loses XP information relative to the LB23 stellar independent model.}

Analogously, in Figure \ref{fig:zgr_cat_wl_errs}, we compare the relative wavelength error distributions for our model to ZGR23 across 10 wavelengths uniformly distributed throughout the XP spectra wavelength range, Eq. \eqref{eq:lam_rel_err}. Then, in Figure \ref{fig:compare_zgr}, we compare the ratio of wavelength reconstruction errors for ZGR23 relative to our model,Eq. \eqref{eq:rmse_lam}.  Here, we only compare errors for stars in the full catalog which satisfy the ZGR23 reliability cut, which decreases the test sample from 50,232 stars down to 37,302 stars. Our stellar label independent model produces less bias from 600-800 nm (in the red half of the XP spectra). Both the wavelength reconstruction errors and relative wavelength error distributions suggest that our model better reconstructs XP spectra for most wavelengths beyond the blue end, relative to ZGR23. 
%We suspect that this is partially explained by ZGR23 explicitly accounting for reddening, which is most relevant for the bluest wavelengths. 
By inspecting the model wavelength error ratios in Figure \ref{fig:compare_zgr}, we observe that our model outperforms ZGR23 in flux reconstruction for most of the wavelengths from 550-900 nm by approximately 20-40\%. However, for wavelengths at the extremities ZGR23 achieves equivalent or better performance than our \emph{s}VAE, concentrated towards the blue end ($<500$ nm). \textcolor{black}{Here, it is informative to compare the APOGEE Kiel diagram in Figure \ref{fig:kiel} to the training sample of ZGR23 \citep[Figure 1 in][]{Zhang++23}. We speculate that the larger number of hot stars in the ZGR23 training sample may contribute to our \emph{s}VAE under-performing at bluer wavelengths.} Lastly, there are certain anomalous wavelengths which do not appear to follow the general error ratio trend as a function of wavelength.

\begin{figure*}
    \centering
    \includegraphics[width=\textwidth]{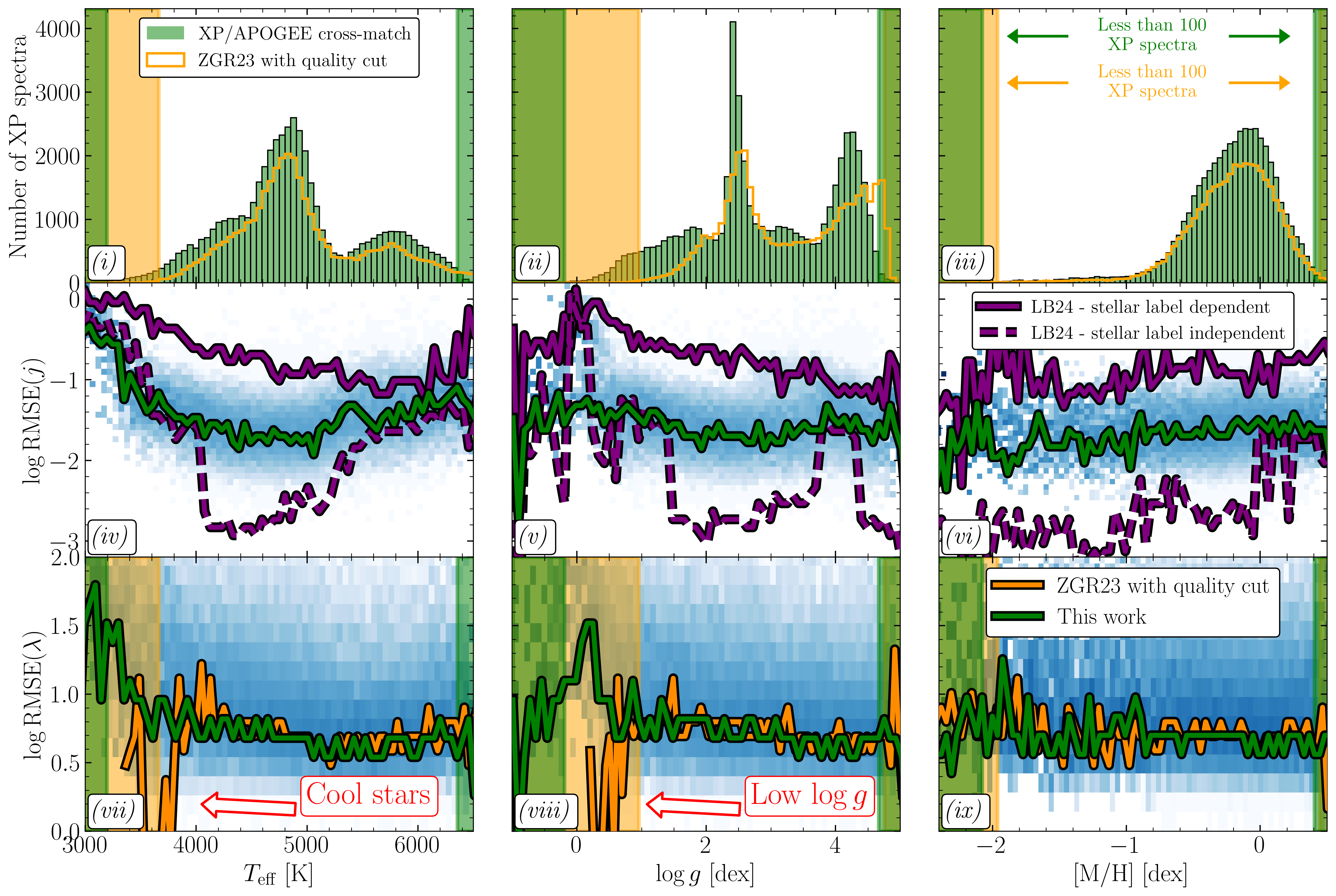}
    \caption{XP spectrum reconstruction errors for LB3 (middle row: \textcolor{black}{\emph{(iv)}-\emph{(vi)}}, purple, coefficient space) and ZGR23 with their quality cut (bottom row: \textcolor{black}{\emph{(vii)}-\emph{(ix)}}, orange, wavelength space) in comparison to our stellar label independent model (green), as a function of stellar labels (top row: \textcolor{black}{\emph{(i)}-\emph{(iii)}}) over test data in the full catalog. Here, we do not apply the ZGR23 quality cut to our wavelength reconstruction errors. Each model curve traces the median reconstruction error as a function of stellar labels. The blue colormap depicts ouur model error distributions, on log scale. Regions of stellar label space with less than 100 XP spectra are shaded in green (orange) for the test spectra (ZGR23 cross-match). Within these regions, stellar label error trends are unreliable. Our stellar label independent model an accommodate both cool \textcolor{black}{(\emph{vii})} and low surface gravity \textcolor{black}{(\emph{viii})} stars, in contrast to the stellar label dependent models (see Section \ref{res:perf} for further discussion).}
    \label{fig:err_labels}
\end{figure*}

Finally, \textcolor{black}{in Figure \ref{fig:err_labels}} we assess our model performance relative to LB23 and ZGR23 over test data in the full catalog, as a function of stellar labels. Importantly, here we do \emph{not} apply the reliability cut of ZGR23 to our model error distributions, but rather only to ZGR23. Since our model is fully independent of stellar labels, we would expect to only observe error trends coming from training object density \textcolor{black}{(panels \emph{(i)}-\emph{(iii)})}, as opposed to genuine trends coming from the labels themselves, which is indeed the case. Our model errors increase towards the wings of the stellar label distributions \textcolor{black}{(panels \emph{(iv)}-\emph{(ix)})} as the number of training objects decreases. First, our XP reconstruction errors in coefficient space are typically 0.5-1 orders of magnitude smaller than the stellar label dependent LB23 model across stellar label space, until training object density decays at the extremities. Conversely, the stellar label independent LB23 model outperforms our model across much of stellar label space. Second, we observe that our model is particularly more robust when applied to cool stars \textcolor{black}{(panels \emph{(iv)} and \emph{(vii)})} and low surface gravity stars \textcolor{black}{(panels \emph{(v)} and \emph{(viii)})}. The ZGR23 error distributions below $T_{\rm eff}\approx4000$ K and below $\log g\approx 1$ dex are effectively meaningless because almost all stars in these regimes are removed by their quality cut. This can be observed by comparing the ZGR23 stellar label distributions (orange) to the XP/APOGEE cross-match stellar label distributions (green). We emphasize that this is an important advantage of our stellar label independent approach: not relying on stellar labels to simulate spectra allows our model to extend into regions of stellar parameter space where labels are unreliable. As such, with an appropriate training sample our model has the potential accurately reproduce both M-dwarf (low temperature) and giant (low surface gravity) BP/RP spectra.

To summarize, our stellar label independent model outperforms the stellar label dependent transformer-based model of LB23, but underperforms in comparison to the stellar label independent LB23 model due to our comparatively harsh data compression. We also compare our model to an ‘expert’ stellar label dependent model: the deep stellar label model of ZGR23. Relative to ZGR23, we find that our stellar label independent model has specific advantages. Namely, simulating XP spectra from approximately 550-990 nm, and simulating cool stars below $T_{\rm eff}\approx4000$ K and low surface gravity stars below $\log g\approx 1$ dex. 

% FIGURE CAPTIONS HAVE BEEN EDITED UP TO AND INCLUDING FIGURE 9
\subsection{Latent space vs. stellar label space}\label{res:latent}

The major novelty of our \emph{s}VAE model is its independence from stellar labels, opting instead to generate an XP spectrum from a latent space. This novelty is a strength, but also a potential weakness. One the one hand, in Section \ref{sec:res}, we demonstrated that our stellar label independent model demonstrates increased performance relative to stellar label dependent models. On the other hand, stellar label dependent models are inherently useful because stellar labels are grounded in astrophysical understanding. Two questions then naturally arise: ‘What astrophysical information has the latent space learned?’ and ‘Is the latent space representation of XP spectra useful?’. We will set aside the latter question and return to it in Sections \ref{sec:app} and \ref{sec:conc}, focusing for now on astrophysical interpretation of the latent space.

\subsubsection{Kiel tracks}
\label{ssec:tracks}

\begin{figure*}
    \centering
    \includegraphics[width=\textwidth]{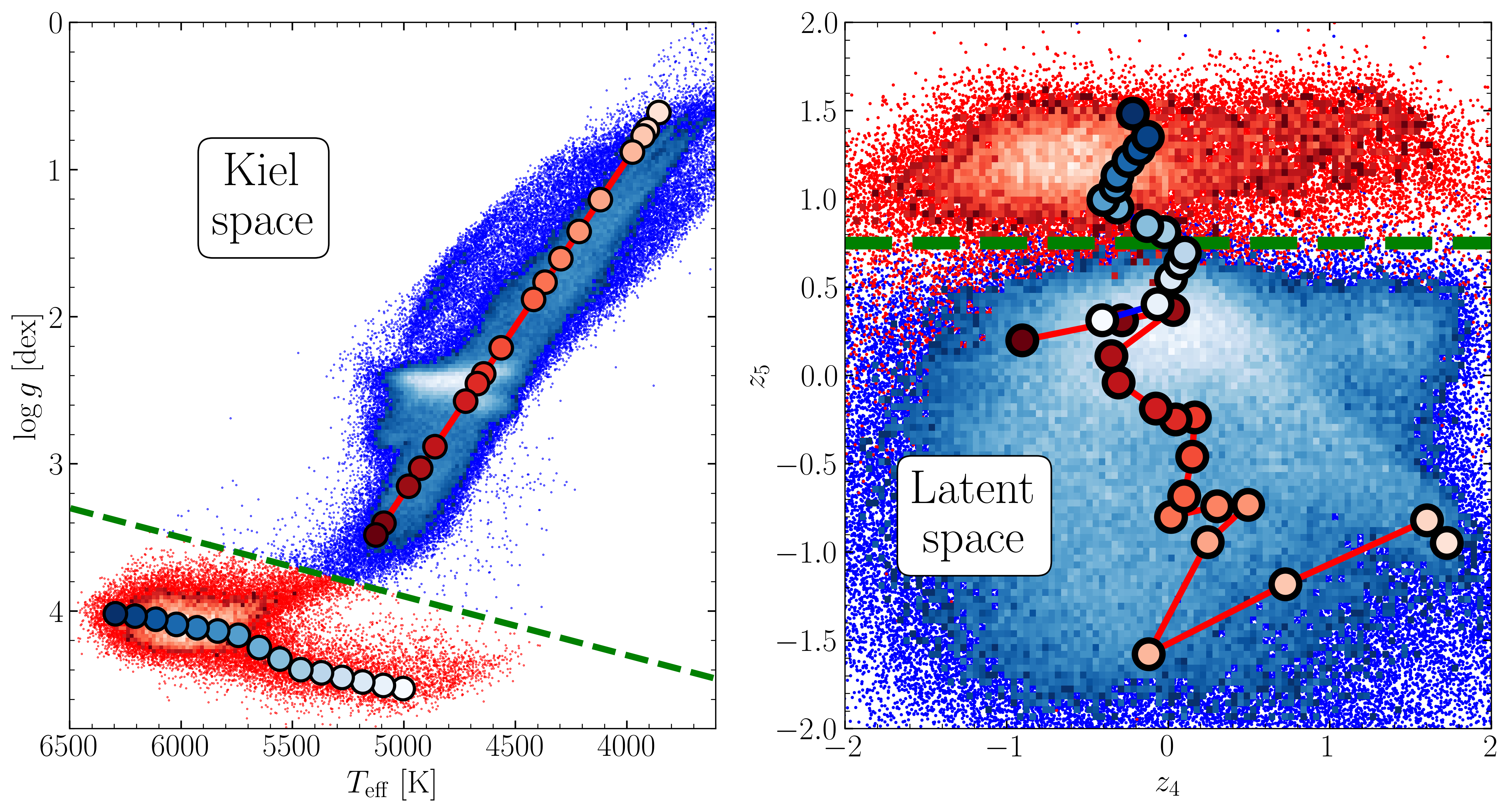}
    \caption{Main-sequence (MS, blue track) and giant branch (GB, red track) evolutionary tracks in Kiel space (left) and in our stellar label independent latent space (right). Additionally, approximate MS/GB (red color map, blue color map) boundaries are presented in both Kiel space and latent space (green dashed lines). Our stellar label independent latent space has learned the distinction between MS and GB stars, as well as the $T_{\rm eff}-\log g$ relation, directly from the data itself.}
    \label{fig:tracks}
\end{figure*}

\begin{figure*}
    \centering
    \includegraphics[width=\textwidth]{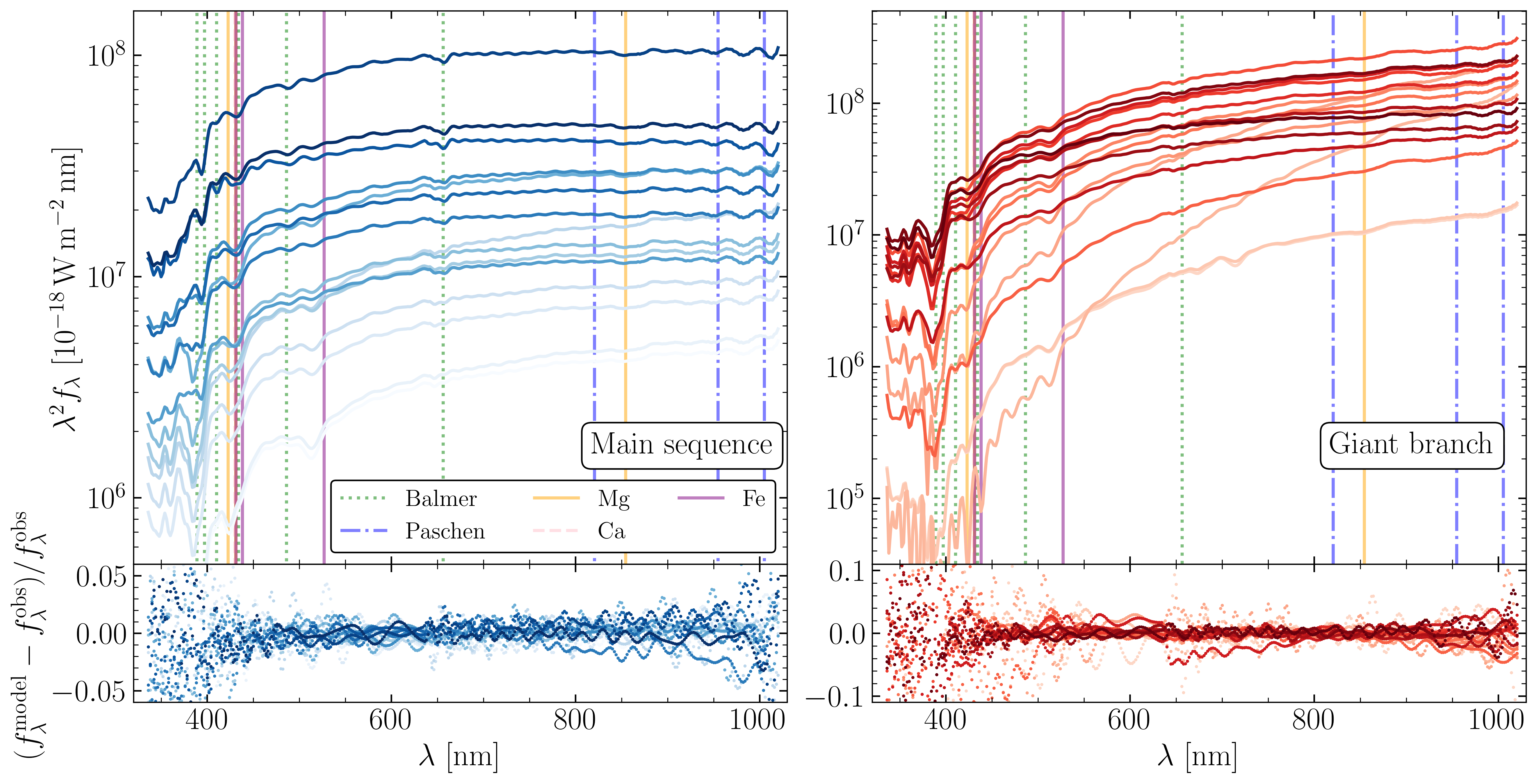}
    \caption{\textcolor{black}{Model-predicted fluxes (at 1 kpc) for representative XP spectra, after de-normalization and scaling by \emph{Gaia G}-flux, along the MS (left) and GB (right) tracks presented in Figure \ref{fig:tracks}, along with percent difference relative to observations (bottom row). The Balmer and Paschen series, as well as Mg, Ca and Fe lines are shown for context.}}
    \label{fig:phys_spectra}
\end{figure*}

We will now argue that our trained \emph{s}VAE learns a latent representation of the XP spectra which contains astrophysical information, using the good labels catalog.

First, we implement an approximate main-sequence/giant branch (MS/GB) cut in label (Kiel) space:
\begin{equation}
    \log g = 5.9 - 0.4\left(\frac{T_{\rm eff}}{k\mathrm{K}}\right),
\end{equation}
visualized as the dashed green line in the left panel of Figure \ref{fig:tracks},
to separate MS and GB stars. In our latent space (right panel of Figure \ref{fig:tracks}), we observe that the $z_5$ latent dimension is effectively a MS/GB classifier: $z_5\approx0.75$ acts as an analogous boundary, above and below which the \emph{s}VAE clusters MS and GB stars, respectively (dashed green line in the right panel of Fig \ref{fig:tracks}). 

Second, we produce MS and GB tracks in $T_{\rm eff}-\log g$ (Kiel) space, at solar-metallicity. We use the Kiel relations which ZGR23, and subsequently LB23, adopted. For MS stars: 
\begin{align}\label{eqn:dwarf}
    \log g_{\rm MS} = 
    \begin{cases}
        4.6, 
        & T_{\rm eff}/k\mathrm{K} < 5 \\
        7.1 - \left(\frac{T_{\rm eff}}{2\,k\mathrm{K}}\right), 
        & 5 \leq T_{\rm eff}/k\mathrm{K} < 6.3 \\
        3.95,
        & T_{\rm eff}/k\mathrm{K} \geq 6.3,
    \end{cases}
\end{align}
and for giants:
\begin{align}\label{eqn:rgb}
    \frac{T_{\rm eff,GB}}{k\mathrm{K}} =
    \begin{cases}
        3.59 + 0.44\log g & \log g < 3.65, \\
        0.09 + 1.4\log g & \log g \geq 3.65.
    \end{cases}
\end{align}
With Eqs. \eqref{eqn:dwarf} and \eqref{eqn:rgb}, we bin and subsequently average XP spectra according to their APOGEE labels along the MS and GB relations in Kiel space to create tracks, shown in the left panel of Figure \ref{fig:tracks}. We then project these spectra into our \emph{s}VAE latent space, and average again to produce analogous tracks--\emph{not as a function of stellar labels, but as a function of latent variables}, shown in the right panel of Figure \ref{fig:tracks}. Here, we only present latent tracks in two of the most Kiel informative latent dimensions (for tracks across the full latent space, see Figure \ref{fig:full_latent_tracks} in Appendix \ref{app:full_latent}). We observe that $z_4$ and $z_5$ function as pseudo-Kiel space, with both the MS and GB tracks exhibiting well-behaved, ordered trajectories through the latent space. \textcolor{black}{That being said, it is apparent that the GB track is slightly more stochastic than the MS track. We speculate that this is could arise from dust extinction which impacts the optical XP spectra, but not the APOGEE stellar labels derived from near-infrared spectroscopy.} We conclude that our \emph{s}VAE learns the $T_{\rm eff}-\log g$ relation, despite the fact that the model has never ‘seen’ these stellar labels. 

\textcolor{black}{Additionally, in Figure \ref{fig:phys_spectra} we present model reconstructed XP spectra in wavelength space for stars randomly drawn from within each bin used to define the MS and GB tracks presented in Figure \ref{fig:tracks}. The percent difference for our reconstructions relative to observations are predominantly at the percent level, apart from the bluest wavelengths where reconstructions deteriorate somewhat (for an in-depth discussion on model reconstruction errors, see Section \ref{res:perf}).}

\subsubsection{Metallicity tracks}\label{ssec:m_h}

\begin{figure*}[!htb]
    \begin{minipage}{0.32\textwidth}
     \centering
     \includegraphics[width=1\linewidth]{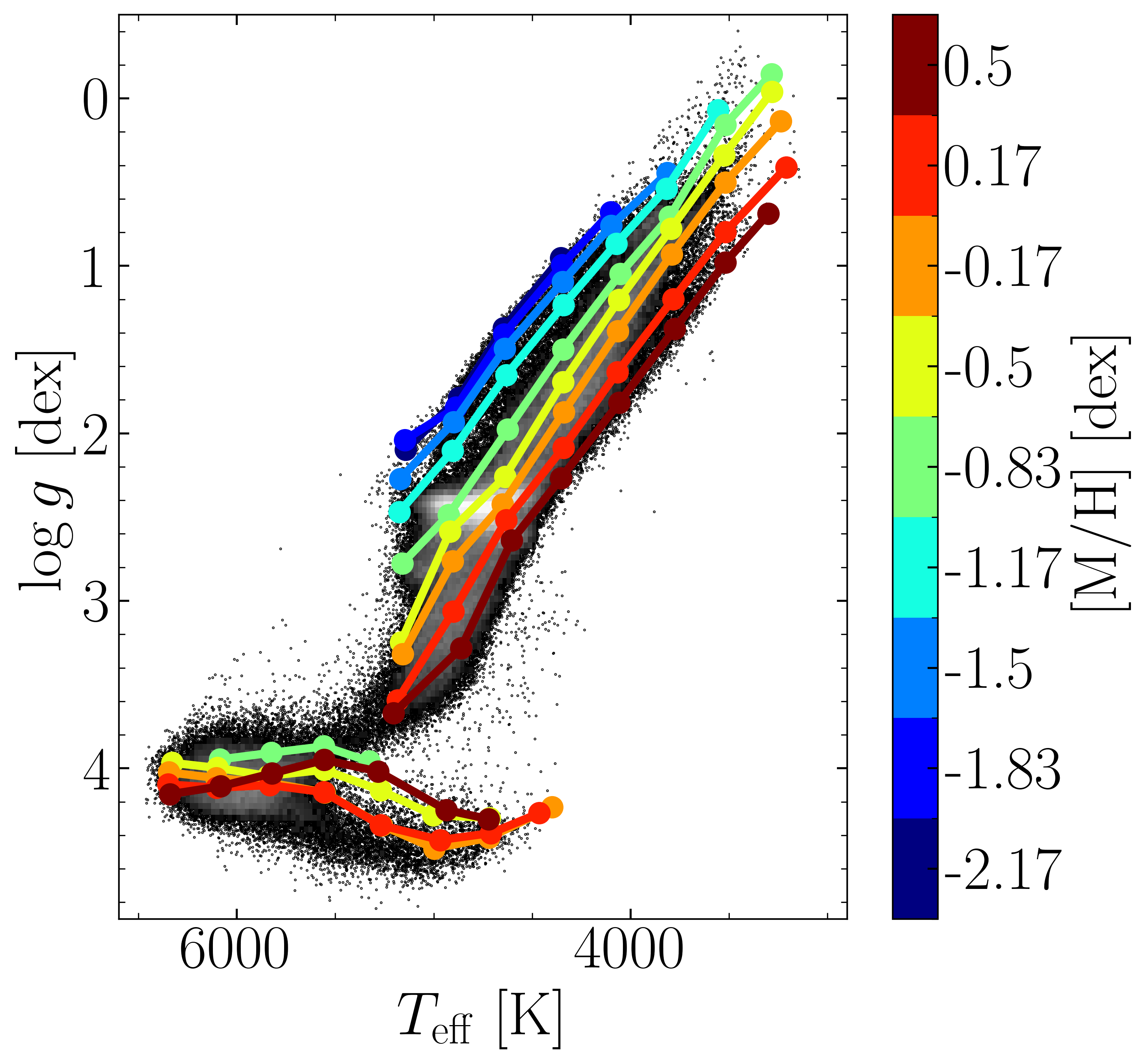}
   \end{minipage}\hfill
   \begin{minipage}{0.32\textwidth}
     \centering
     \includegraphics[width=1\linewidth]{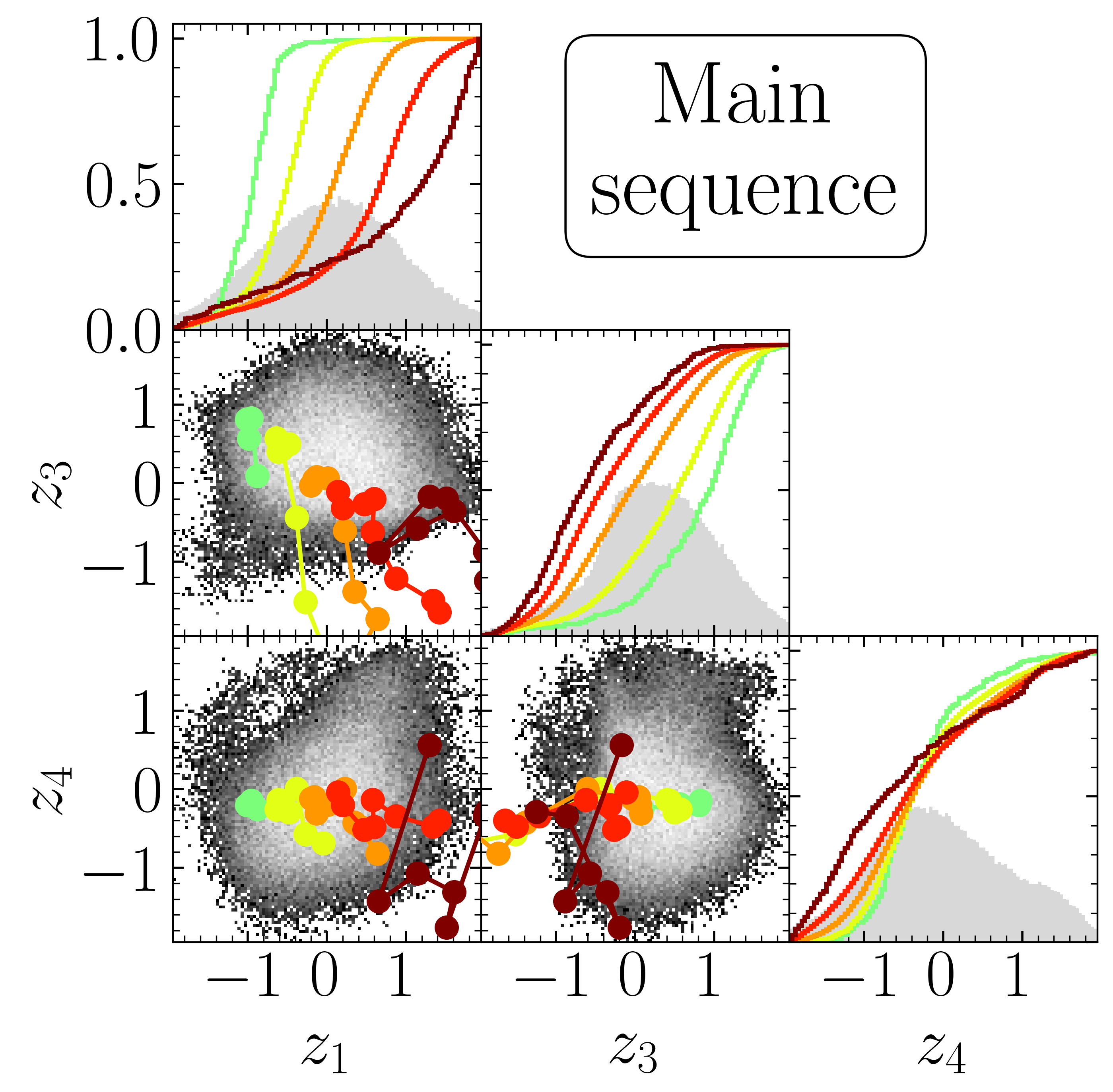}
   \end{minipage}\hfill
   \begin{minipage}{0.32\textwidth}
     \centering
     \includegraphics[width=1\linewidth]{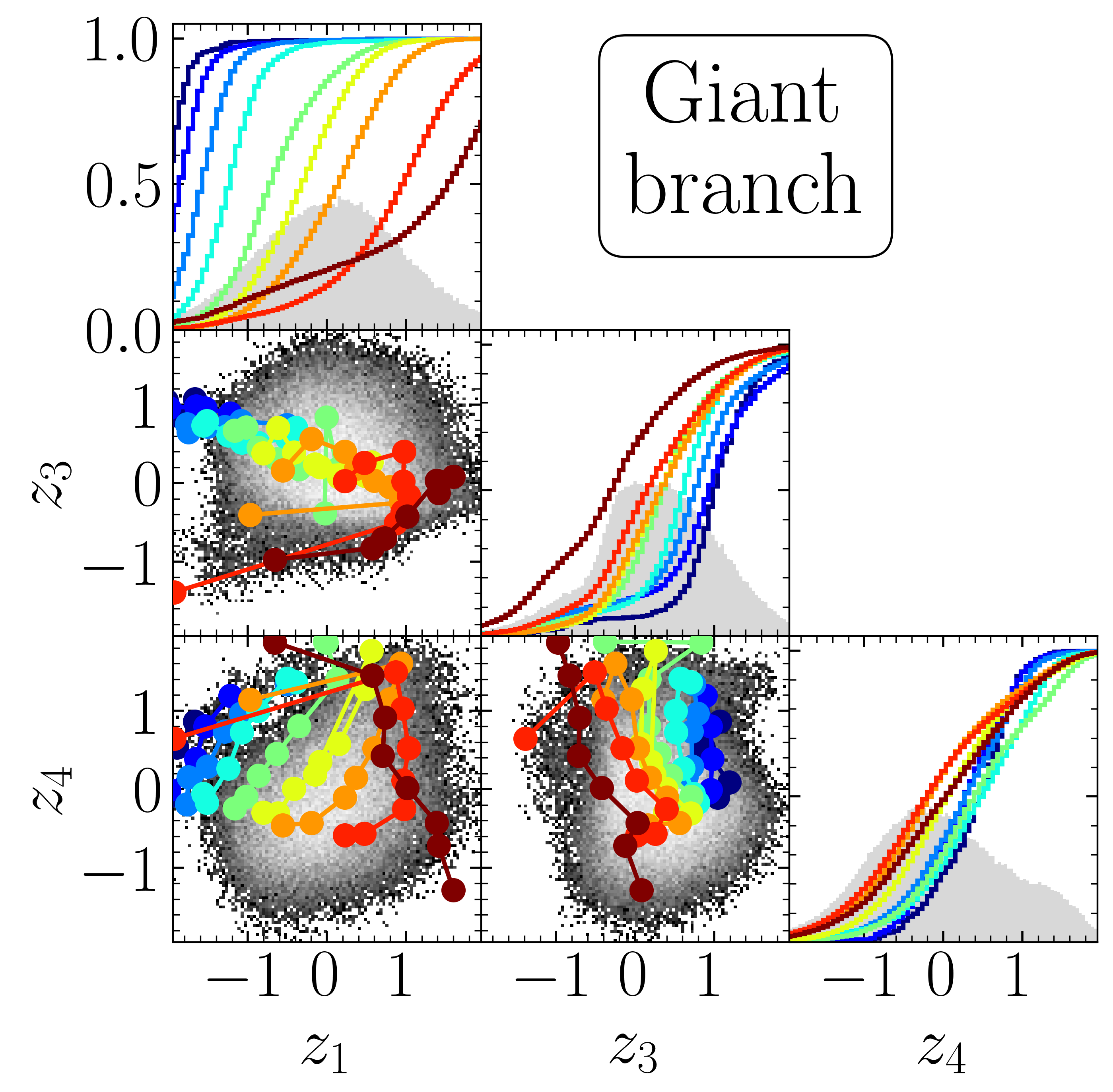}
   \end{minipage}
   \caption{Fixed metallicity tracks along the MS and GB in Kiel space (left), projected into the latent space (MS middle, GB right). The three most metal informative latent dimensions are presented. In  the latent space 1D marginal distributions, cumulative distribution at fixed metallicity are also presented. Metallicity gradients across the latent dimensions demonstrate that our stellar label dependent model has learned [M/H] information directly from the data itself.}
   \label{fig:metal_tracks}
\end{figure*}

We have established that our \emph{s}VAE learns classification of MS and giant stars and stellar evolutionary tracks along the MS and GB. But what about metallicity? Analogously to the tracks in Sect. \ref{ssec:tracks}, we first produce tracks at fixed metallicity in $T_{\rm eff}-\log g$ space, and subsequently translate them into our \emph{s}VAE latent space.
In the left panel of Figure \ref{fig:metal_tracks}, we present binned Kiel tracks along both the MS and GB from high ($\rm[M/H]\approx0.5$) to low ($\rm[M/H]\approx-2.2$) metallicity. Then, in the middle (right) panel of Figure \ref{fig:metal_tracks} we project the MS (GB) metallicity tracks into our \emph{s}VAE latent space, and present the three most metal-informative latent dimensions.

We observe evident metallicity gradients in the latent space, for both MS and giant stars. This is demonstrated by both the clear separation of fixed metallicity tracks in both the 2D latent space distributions and 1D cumulative distributions. Furthermore, $z_1$ is the most metallicity dominated latent dimension, with a strong metallicity gradient from metal-poor stars at $z_1\ll0$ to metal-rich stars at $z_1\gg0.$ Interestingly, the main sequence metallicity tracks, which are crowded and overlapping in Kiel space, appear to be better separated in the latent space (along $z_1$).

In summary, our trained \emph{s}VAE learns information about \emph{Gaia} XP spectra which is effectively equivalent to stellar labels: $T_{\rm eff},\,\log g$ and $\rm [M/H],$ without ever having ‘seen’ any of these labels during training. However, the latent space blends its understanding of stellar labels across multiple latent dimensions. The information our \emph{s}VAE learns about \emph{Gaia} XP spectra contains stellar label information, but can additionally learn information from stellar spectra which is not captured by stellar labels. Furthermore, our latent space can be used to probe the actual astrophysical information content of the \emph{Gaia} XP spectra, without the ambiguity produced by stellar label correlations; an important consideration for $\alpha$-abundance.

\section[f]{[$\alpha$/M] information in the Gaia XP Spectra}\label{sec:app}

The chemical distribution of stars in the Galactic disk of our Milky Way contains important information about the formation, accretion and dynamical evolutionary histories of the Galactic disk. In particular, observations of Galactic disk stars have long been known to exhibit two components in [M/H]-[$\alpha$/M] space: the $\alpha$-bimodality, comprised of the high-$\alpha$ and low-$\alpha$ sequences \citep{Fuhrmann98,Bensby++03,Anders++14,Nidever++14,Hayden++15,Kordopatis++15}. Initially, these sequences were used to separate the thin and thick disks \citep{Yoshii82,GilmoreReid83}. The thick (thin) disk is mostly comprised of the high-$\alpha$ (low-$\alpha$) sequence, made up of old, kinematically hot (young, kinematically cold) stars. More recently, this simple picture of geometric distinction between $\alpha$-sequences has been disfavored \citep{SchonrichBinney09,Bovy++12a,Bovy++12c,Bovy++12b,Bland-Hawthorn++19}. However, it is nevertheless clear that there are two distinct $\alpha$-sequences, now thought to be the result of chemical enrichment  \citep{Hayden++17}.

The estimation of $\alpha$-abundance from \emph{Gaia} XP spectra has recently been explored in the literature. In particular, several stellar label dependent analyses reach different conclusions: 

Initially, \cite{Gavel++21} used the \texttt{ExtraTrees} algorithm and argued that XP spectra $\alpha$-abundance measurements for stars which \emph{only} vary in [$\alpha$/M] are dangerous because XP $\alpha$-abundance estimates strongly depend on correlations between [$\alpha$/M] and other stellar properties, particularly [M/H]. This is potentially because the \emph{Gaia} XP spectra are so extremely low-resolution that a change in [$\alpha$/M] is equivalent to a change in [M/H] due to multiple spectral lines being blurred together. Then, \cite{Witten++22} supported these findings by demonstrating that synthetic XP spectra do not contain significant $\alpha$-information. It is important to mention that both of these works emphasize that these difficulties are especially relevant for warm stars with 5000 K $> T_{\rm eff}.$ Additionally, the work of \cite{Witten++22} involved synthetic XP spectra with $G=16.$

More recently, \cite{ASPGAP-23} developed \texttt{AspGap}: a regression model which they argue is capable of producing precise [$\alpha$/M] estimates for red giant branch stars with $3000 \leq T_{\rm eff} \leq 7000$ K, which extends into the warm regime where $\alpha$-abundance estimates have been identified as problematic. Then, \cite{Hattori++24} trained a tree-based machine learning model to estimate both [M/H] and [$\alpha$/M]. Additionally, \cite{Hattori++24} proposed specific lines which contribute to $\alpha$-information: Na D lines (589 nm) and the Mg I line (516 nm).

We emphasize that the conclusions drawn by these previous analyses about $\alpha$-information in \emph{Gaia} XP are obfuscated by their reliance on stellar labels. Indirect stellar label correlations with [$\alpha$/M] cannot be separated from a stellar label dependent model. On the other hand, our stellar label independent model can potentially assess $\alpha$-information without the issue of stellar label correlations.

\begin{figure*}[!htb]
     \centering
     \includegraphics[width=\textwidth]{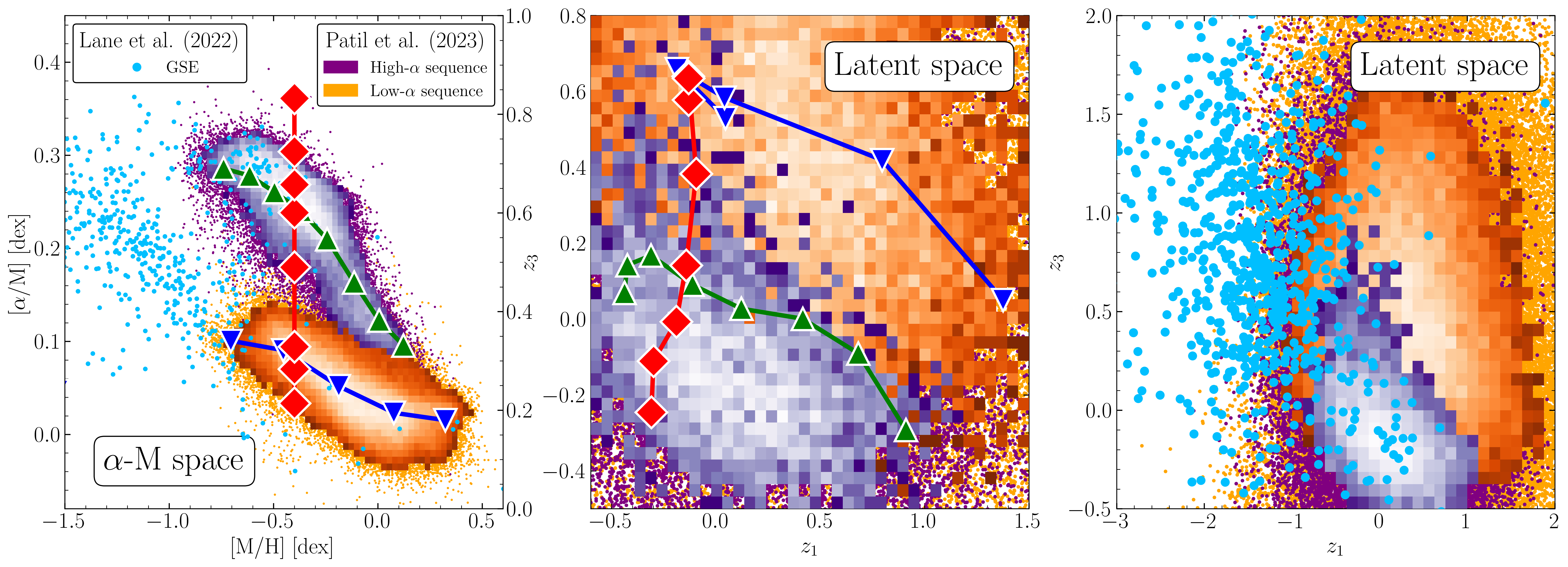}
     \caption{High-$\alpha$ sequence (green), low-$\alpha$ sequence (blue) and fixed-metallicity (red) tracks in both [$\alpha$/M]-[M/H] space (left) and latent space (center). High/low $\alpha$-sequence members (purple, orange) are classified according to the $\alpha$-bimodality split of \cite{Patil++23}. $\alpha$-information is demonstrated by the high- and low-$\alpha$ sequence tracks, in combination with the fixed metallicity track (which only varies in [$\alpha$/M]). \textcolor{black}{Additionally, we present a sample of \emph{Gaia}-Enceladus stars selected via action diamond cuts from \cite{Lane+22} (light blue) in both [$\alpha$/M]-[M/H] space (left) and latent space (right).}}
     \label{fig:alpha_bi}
\end{figure*}

To contribute to answering the question of $\alpha$-information in \emph{Gaia} XP, we separate high- and low-$\alpha$ sequence members of the pristine labels catalog\footnote{The pristine labels catalog is strictly composed of giants with $T_{\rm eff} < 5000$ K. As such, we are probing the cool star regime where $\alpha$-abundance estimates are less problematic.} in [M/H]-[$\alpha$/M] space (left panel of Figure \ref{fig:alpha_bi}) and project them into our latent space (middle panel of Figure \ref{fig:alpha_bi}). To do so, we implement the $\alpha$-bimodality split of \cite{Patil++23}, who used copulas and elicitable maps to cleanly separate the high- (purple) and low-$\alpha$ (orange) sequences\footnote{\cite{Patil++23} define their split in [Fe/H]-[Mg/Fe] space. As such, we do the same, and subsequently present the split in [M/H]-[$\alpha$/M] space.}. We then produce binned tracks along both sequences in [M/H]-[$\alpha$/M] space and project them into our latent space, similarly to the tracks in Section \ref{sec:res}. We observe that our $z_1$ and $z_3$ latent variables function as pseudo-[M/H]-[$\alpha$/M] space, in which the high- and low-$\alpha$ sequence tracks are clearly separated. However, these tracks alone are not definitive evidence of $\alpha$-information, because they are functions of metallicity! Therefore, we produce a third track at fixed metallicity ([M/H] $=-0.4$, red), which only varies with [$\alpha$/M]. Crucially, the fixed metallicity track transitioning between both sequences in [M/H]-[$\alpha$/M] space is reproduced in our latent space. The high- and low-$\alpha$ sequence tracks, with the addition of the fixed metallicity track, is compelling evidence that the \emph{Gaia} XP spectra contain meaningful $\alpha$-information (for low-temperature stars).

\textcolor{black}{To reinforce our argument that the latent space has learned important chemical information, we also present a sample of \emph{Gaia}-Enceladus stars \citep[e.g.][]{Helmi+18} in both [M/H]-[$\alpha$/M] space (left panel of Figure \ref{fig:alpha_bi}) and latent space (right panel of Figure \ref{fig:alpha_bi}). To select \emph{Gaia}-Enceladus members, we implement the action diamond cuts recommended by \cite{Lane+22}: $|L_z/J_{\rm tot}|<0.07$ and $(J_z-J_R)/J_{\rm tot}<-0.3,$ for stars in the full catalog. The evident separation of the \emph{Gaia}-Enceladus population from both the high- and low-$\alpha$ sequences in the latent space (particularly across $z_1$) demonstrates that our model is sensitive to the anomalous abundances characteristic of \emph{Gaia}-Enceladus.}

\begin{figure*}[!htb]
     \centering
     \includegraphics[width=\textwidth]{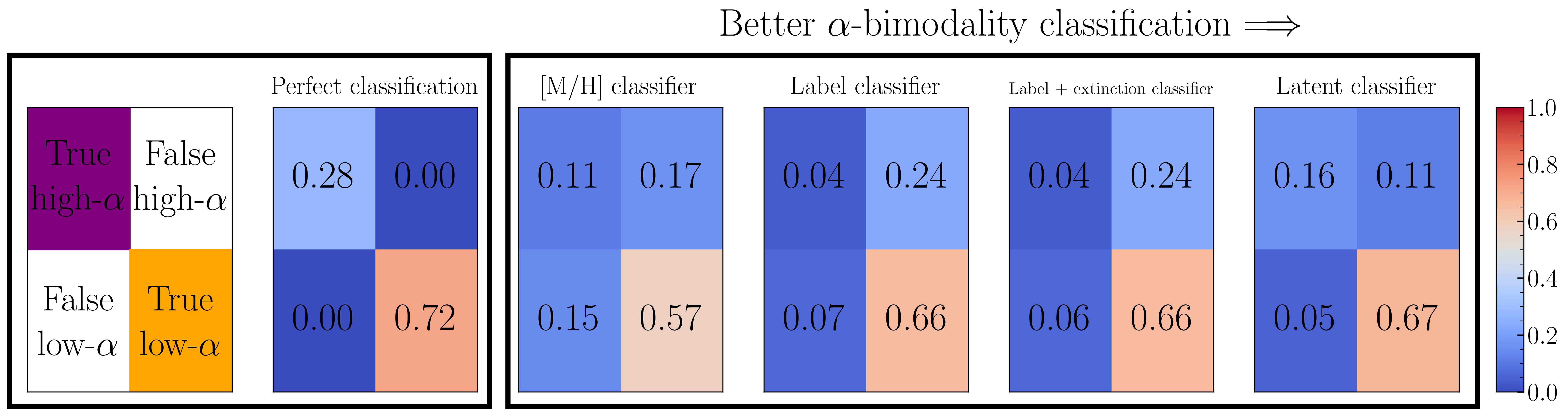}
     \caption{Confusion matrices for the label, [M/H] and latent classifiers presented in Section \ref{sec:app}. Confusion matrix classifications and perfect classification value are presented in the two left-most matrices. The latent classifier, trained on our \emph{s}VAE latent variables, achieves better $\alpha$-bimodality classification than classifiers relying on stellar labels.}\label{fig:cms}
\end{figure*}

To be even more pessimistic, one can also worry about $T_{\rm eff}$ and $\log g$ correlations with [$\alpha$/M] which could masquerade as [$\alpha$/M] trends in our latent space. Furthermore, it is also possible that the underlying extinction distributions for the high-$\alpha$ and low-$\alpha$ populations differ significantly. Therefore, it is important to demonstrate that our model actually learns the $\alpha$-bimodality without relying on stellar label and extinction relationships beyond metallicity. \textcolor{black}{Lastly, some work has already been undertaken to present evidence for $\alpha$-information content in a similar manner to the above. For example, synthetic photometry from Gaia XP spectra has been used to demonstrate that [$\alpha$/M] appears to correlate with the \emph{Gaia} colour C1M395-C1M410 (see Fig. 28 of \citep{2022arXiv220606215G}).}

As such, we train four Random Forest Classifiers \citep[\texttt{sklearn.ensemble.RandomForestClassifer};][]{scikit-learn-11} on 90\% of the pristine catalog to classify the high- and low-$\alpha$ sequences, \textcolor{black}{with the intent of quantitatively demonstrating $\alpha$-information in a novel way. The classifiers are as follows:}
\begin{enumerate}[label=(\roman*)]
    \item \emph{Metal classifier}: trained on only [M/H].
    \item \emph{Label classifier}: trained on all stellar labels \textcolor{black}{except} [$\alpha$/M]: $T_{\rm eff},$ $\log g$ and [M/H].
    \item \emph{Label + extinction classifier}: trained on all stellar labels except [$\alpha$/M], as well as 2MASS colours: $J-H$ and $J-K$, to provide training labels which can proxy extinction.
    \item \emph{Latent classifier}: trained on latent variables from our stellar label independent model.
\end{enumerate}
We present the confusion matrices for the four classifiers in Figure \ref{fig:cms} over test data (10\% of the pristine labels catalog). To compare the classifiers, we use accuracy:
\begin{equation}
    \text{Acc} = \frac{\text{T high-}\alpha + \text{T low-}\alpha}{\text{T high-}\alpha + \text{T low-}\alpha + \text{F high-}\alpha + \text{F low-}\alpha},
\end{equation}
where T is true and F is false. Furthermore, we define
the high-$\alpha$ predictive value:
\begin{equation}
    P_{\rm high-\alpha} = \frac{\text{T high-}\alpha}{\text{T high-}\alpha + \text{F high-}\alpha},
\end{equation}
and the low-$\alpha$ predictive value:
\begin{equation}
    P_{\rm low-\alpha} = \frac{\text{T low-}\alpha}{\text{T low-}\alpha + \text{F low-}\alpha}.
\end{equation}
We present Acc, $P_{\rm high-\alpha}$ and $P_{\rm low-\alpha}$ for each classifier in Table \ref{tab:class_metrics}. Classifying the $\alpha$-bimodality with our stellar label independent latent space yields a 16\% (14\%) improvement in accuracy relative to the [M/H] (labels) classifier. The latent classifier achieves 18\% (14\%) improvement in high-$\alpha$ (low-$\alpha$) predictive value relative to the [M/H] classifier. Interestingly, the labels classifier achieves comparable low-$\alpha$ predictive value to the latent classifier. However, this is at the expense of a far worse high-$\alpha$ predictive value: the latent classifier achieves a 43\% improvement relative to the labels classifier. This is a somewhat counter-intuitive result: including $T_{\rm eff}$ and $\log g$ produces worse high-$\alpha$ classification than [M/H] alone. This is additional evidence that our latent space is learning genuine $\alpha$-information, since the inclusion of stellar labels beyond [M/H] decreases high-$\alpha$ classification performance as opposed to improving high-$\alpha$ classification performance. Finally, the inclusion of 2MASS $J-H$ and $J-K$ colours does not yield any significant classification performance, which indicates that our latent classifier is not relying on different extinction trends between $\alpha$-sequences.

In summary, the striking $\alpha$-bimodality tracks in our latent space, in combination with the significant improvement in $\alpha$-bimodality classification with our latent classifier relative to the [M/H], label and label + extinction classifiers, is stellar label independent evidence that the \emph{Gaia} XP spectra can and should be used to estimate [$\alpha$/M] with $T_{\rm eff} < 5000$ K. \textcolor{black}{We speculate that it may be possible to estimate [$\alpha$/M] above this temperature threshold if a training sample with hotter stars than APOGEE were used.}

\begin{table}
    \caption{Performance metrics for the three $\alpha$-bimodality classifiers presented in Section \ref{sec:app}. A perfect classifier would yield accuracy Acc $=1,$ high-$\alpha$ predictive value $P_{\rm high-\alpha}=1$ and low-$\alpha$ predictive value $P_{\rm low-\alpha}=1.$ }
    \centering
    \begin{tabular}{|c|c|c|c|c|}
        \hline
        & [M/H] & Label & Label + extinction & Latent \\
        \hline
        Acc & 0.68 & 0.70 & 0.70 & 0.84 \\
        $P_{\rm high-\alpha}$ & 0.40 & 0.15 & 0.15 & 0.58 \\
        $P_{\rm low-\alpha}$ & 0.79 & 0.91 & 0.92 & 0.93 \\
        \hline
    \end{tabular}
    \label{tab:class_metrics}
\end{table}

\section{Summary, discussion and outlook}\label{sec:conc}

We have presented a stellar label independent model for the recently released \emph{Gaia} XP spectra from Gaia DR3. The primary goal of this work was to demonstrate that stellar label independent models can close the stellar labels gap, and in doing so rectify systematics which indirectly stem from theoretically estimated stellar labels. Our main results are as follows:
\begin{itemize}
    \item[(i)] \emph{Model performance}: Our stellar label independent model achieves competitive, and in specific regimes better, performance relative to existing stellar label dependent models. Specifically, our \emph{scatter} VAE better reconstructs XP spectra over the redder XP wavelengths. Also, our model can be applied to cool stars and low-surface gravity stars: stellar label regions where stellar label dependent models struggle.
    \item[(ii)] \emph{Interpretability}: We investigated stellar label trends in our stellar label independent latent space to demonstrate that the latent space is rich with astrophysical information, such as: main-sequence and giant branch evolution, and metallicity.
    \item[(iii)] [$\alpha$/M] \emph{information}: We provided strong evidence that the \emph{Gaia} XP spectra contain meaningful $\alpha$-abundance information without relying on stellar label correlations, for stars with $T_{\rm eff}<5000$, \textcolor{black}{while also being sensitive to the anomalous abundances of \emph{Gaia}-Enceladus stars.} As such, we encourage the astronomy community to make use of the \emph{Gaia} XP spectra for estimation of [$\alpha$/M] in this temperature regime\footnote{The extension of [$\alpha$/M] estimates to hotter stars in the \emph{Gaia} XP spectra was not addressed in this work.}.
\end{itemize}
Despite the successes of our stellar label independent approach, our model has some limitations. Specifically:
\begin{itemize}
    \item[(i)] \emph{Indirect astrophysical information}: By definition, our stellar label independent model does not produce stellar label estimates. As such, extracting astrophysical information from our model is not necessarily straightforward. Nevertheless, our model can provide important astrophysical understanding by investigating the latent space clustering of XP spectra.
    \item[(ii)] \emph{Mileage may vary}: The variational nature of a VAE, implemented with the re-parametrization trick, means that each projection of an XP spectrum into the latent space, or simulation of an XP spectrum from the latent space, is not unique. Results and figures in this work may vary slightly if recomputed as a result. However, these variations should not be severe, since the latent variable sampling is restricted to $\epsilon$ (see Section \ref{sec:meth}).
    \item[(iii)] \textcolor{black}{\emph{Gaia XP spectra coverage}}: The first implementation of our stellar label independent model was trained on a small minority of XP spectra relative to the total number of XP spectra available ($<1\%$). As such, we caution the application of our model to stellar populations beyond the XP/APOGEE cross-match, such as white dwarfs. This does not mean that our model cannot eventually accommodate these populations. Future implementations of our stellar label \textcolor{black}{independent} model can be trained on any/all XP spectra because our model does not require stellar labels to train on. 
\end{itemize}
The potential of our stellar label independent approach was certainly not fully explored in this initial work. We identify several promising areas for future work:
\begin{itemize}
    \item[(i)] \emph{Denoising}: Data compression into the latent space can remove noise from input XP spectra, because our model is designed to learn the most important features shared by the training data and neglect observational noise. Imposing restrictions on our scatter estimator could produce de-noised XP spectra with model error estimates smaller than observational uncertainties.
    \item[(ii)] \emph{Global XP model}: In principle, our stellar label independent model could be trained to accurately simulate the entire XP dataset. This would require a sophisticated training procedure due to the gigantic amount of training data required for a single model to learn all of the stellar populations in XP spectra ($\sim10^8$ training objects).
    \item[(iii)] \emph{Outlier detection}: A particularly promising application of our stellar label independent model is the detection of rare sub-populations in the \emph{Gaia} XP spectra via the latent space. Recently, \cite{Lucey++22} released a catalog of carbon-enhanced metal poor stars candidates (CEMPs). A preliminary application of our model to their CEMP catalog has uncovered a population of outliers. We suspect that a fraction of these outliers could be binary systems containing CEMP stars, since a major formation channel for CEMPs is binary mass transfer \citep[e.g.][]{Goswami++21}. We will soon present a detailed analysis of this outlier population in \emph{Laroche et al. 2025 (in prep.)}.
\end{itemize}

In conclusion, the \emph{Gaia} DR3 XP spectra is by far the largest spectroscopic survey to date. Stellar label dependent models are incapable of exploiting the entirety of the astrophysical information in the XP dataset, because they are limited by the availability of stellar labels to train on. Novel data-driven techniques must be developed to tackle this big data problem. Furthermore, as the sizes of future large-scale surveys grow, the stellar labels gap for stellar label dependent models will only widen as the relative stellar label coverage decreases (e.g. SPHEREx: \citealp{SPHEREx-16}; Roman: \citealp{Roman22}; Vera Rubin: \citealp{LSST-19}). The big-data era in astronomy is both a blessing, for discovery, and a curse, for data analysis. Our stellar label independent model is an important step towards \emph{fully} data-driven modeling which can confront the intimidating amount of data which present and future large-scale surveys provide.

\section*{Acknowledgments}
\textcolor{black}{We are thankful to the anonymous referee for providing a constructive report that helped improve the clarity and quality of our manuscript.} AL and JSS would like to thank Maria Drout, Jo Bovy, Henry Leung, Gregory M. Green, Jiadong Li and Jeff Shen for valuable feedback. AL also acknowledges helpful conversations with James Lane, Nolan Koblischke and Amy Prickett which greatly improved this work. AL acknowledges support from the Natural Sciences and Engineering Research Council of Canada (NSERC) and is (was) partially funded through a NSERC Canada Graduate Scholarship - Doctoral (Master’s). AL is also supported by the Data Sciences Institute at the University of Toronto through grant number DSI- DSFY3R1P02. JSS acknowledges support from the Natural Sciences and Engineering Research Council of Canada (NSERC) funding reference \#RGPIN-2023-04849. The Dunlap Institute is funded through an endowment established by the David Dunlap family and the University of Toronto. 

This work has made use of data from the European Space Agency (ESA) mission
{\it Gaia} (\url{https://www.cosmos.esa.int/gaia}), processed by the {\it Gaia}
Data Processing and Analysis Consortium (DPAC,
\url{https://www.cosmos.esa.int/web/gaia/dpac/consortium}). Funding for the DPAC
has been provided by national institutions, in particular the institutions
participating in the {\it Gaia} Multilateral Agreement. Additionally, this work made use of the Python package \texttt{GaiaXPy}, developed and maintained by members of the Gaia DPAC, and in particular, Coordination Unit 5 (CU5), and the Data Processing Centre located at the Institute of Astronomy, Cambridge, UK (DPCI).

\appendix

\section{Complete latent space representation}\label{app:full_latent}

\begin{figure}
    \centering
    \includegraphics[width=\textwidth]{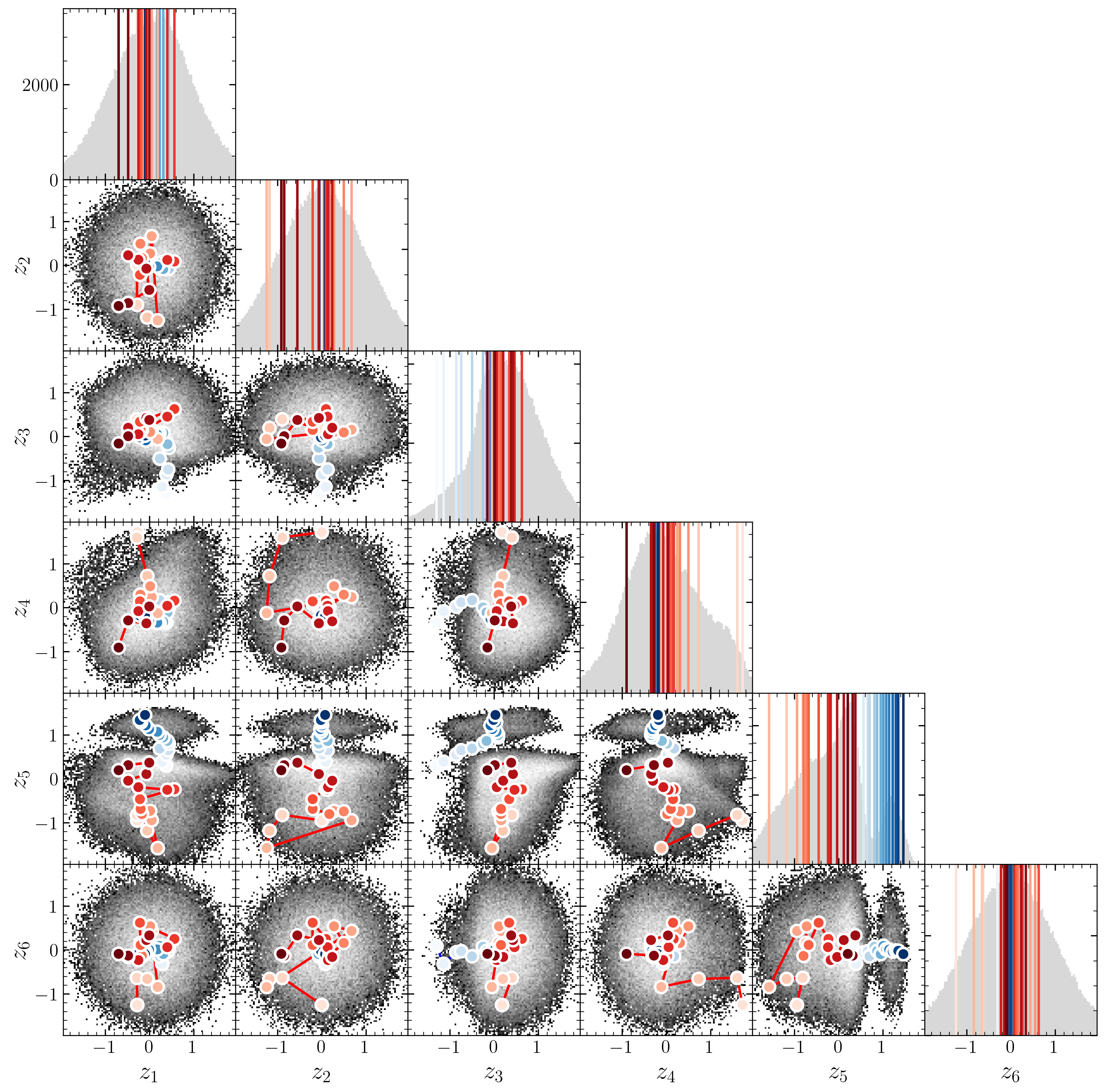}
    \caption{Main-sequence (MS, blue track) and giant branch (GB, red track) evolutionary tracks as a function all latent variables (the same tracks presented in Figure \ref{fig:kiel}). 1D and 2D marginal distributions are presented over the entire 6D latent space. In the 1D marginal distributions, evolutionary track points are presented as vertical lines. Stellar label information is shared across all latent dimensions.}
    \label{fig:full_latent_tracks}
\end{figure}

By inspecting the latent tracks across all six dimensions in Figure \ref{fig:full_latent_tracks}, it is clear that the Kiel information our model has learned is not exclusively encoded into $z_4$ and $z_5$ (see Figure \ref{fig:kiel}). Rather, the stellar label information is shared amongst all latent dimensions, with varying degrees of sensitivity to stellar labels. This additional level of complexity, relative to stellar label dependent models which do not blend stellar label information across summary statistics, can lead to better performance (see Section \ref{res:perf}). Simultaneously, the blending of astrophysical information across the latent space makes the assessment of our model behaviour less straightforward. This also means that our latent dimensions can learn information beyond stellar labels. As such, our stellar label independent framework is very much optimized for outlier detection.

\bibliography{sample631}{}
\bibliographystyle{aasjournal}

\end{document}